\documentclass[preprint,superscriptaddress,amsmath,amssymb,aps,prl]{revtex4-1}
\usepackage{graphicx,epstopdf}% Include figure files
\usepackage{dcolumn}% Align table columns on decimal point
\usepackage{bm}% bold math
\usepackage{hyperref}% add hypertext capabilities
\usepackage{multirow}
\usepackage{threeparttable}
\usepackage{color}
\usepackage{subfigure}
\usepackage{setspace}
\usepackage{mathtools}
\usepackage{tipa}

\newcommand{\ket}[1]{\ensuremath{\left|#1\right\rangle}}
\newcommand{\bra}[1]{\ensuremath{\left\langle#1\right|}}

\begin{document}

%\preprint{APS/123-QED}

\title{Measuring the winding number in a large-scale chiral quantum walk}% Force line breaks with \\
%\thanks{A footnote to the article title}%

\author{Xiao-Ye Xu}
\affiliation{CAS Key Laboratory of Quantum Information, University of Science and Technology of China, Hefei 230026, People's Republic of China}
\affiliation{Synergetic Innovation Center of Quantum Information and Quantum Physics, University of Science and Technology of China, Hefei 230026, People's Republic of China}
\author{Qin-Qin Wang}
\affiliation{CAS Key Laboratory of Quantum Information, University of Science and Technology of China, Hefei 230026, People's Republic of China}
\affiliation{Synergetic Innovation Center of Quantum Information and Quantum Physics, University of Science and Technology of China, Hefei 230026, People's Republic of China}
\author{Wei-Wei Pan}
\affiliation{CAS Key Laboratory of Quantum Information, University of Science and Technology of China, Hefei 230026, People's Republic of China}
\affiliation{Synergetic Innovation Center of Quantum Information and Quantum Physics, University of Science and Technology of China, Hefei 230026, People's Republic of China}
\author{Kai Sun}
\affiliation{CAS Key Laboratory of Quantum Information, University of Science and Technology of China, Hefei 230026, People's Republic of China}
\affiliation{Synergetic Innovation Center of Quantum Information and Quantum Physics, University of Science and Technology of China, Hefei 230026, People's Republic of China}
\author{Jin-Shi Xu}
\affiliation{CAS Key Laboratory of Quantum Information, University of Science and Technology of China, Hefei 230026, People's Republic of China}
\affiliation{Synergetic Innovation Center of Quantum Information and Quantum Physics, University of Science and Technology of China, Hefei 230026, People's Republic of China}
\author{Geng Chen}
\affiliation{CAS Key Laboratory of Quantum Information, University of Science and Technology of China, Hefei 230026, People's Republic of China}
\affiliation{Synergetic Innovation Center of Quantum Information and Quantum Physics, University of Science and Technology of China, Hefei 230026, People's Republic of China}
\author{Jian-Shun Tang}
\affiliation{CAS Key Laboratory of Quantum Information, University of Science and Technology of China, Hefei 230026, People's Republic of China}
\affiliation{Synergetic Innovation Center of Quantum Information and Quantum Physics, University of Science and Technology of China, Hefei 230026, People's Republic of China}
\author{Ming Gong}
\affiliation{CAS Key Laboratory of Quantum Information, University of Science and Technology of China, Hefei 230026, People's Republic of China}
\affiliation{Synergetic Innovation Center of Quantum Information and Quantum Physics, University of Science and Technology of China, Hefei 230026, People's Republic of China}
\author{Yong-Jian Han}
\email{smhan@ustc.edu.cn}
\affiliation{CAS Key Laboratory of Quantum Information, University of Science and Technology of China, Hefei 230026, People's Republic of China}
\affiliation{Synergetic Innovation Center of Quantum Information and Quantum Physics, University of Science and Technology of China, Hefei 230026, People's Republic of China}
\author{Chuan-Feng Li}
\email{cfli@ustc.edu.cn}
\affiliation{CAS Key Laboratory of Quantum Information, University of Science and Technology of China, Hefei 230026, People's Republic of China}
\affiliation{Synergetic Innovation Center of Quantum Information and Quantum Physics, University of Science and Technology of China, Hefei 230026, People's Republic of China}
\author{Guang-Can Guo}
\affiliation{CAS Key Laboratory of Quantum Information, University of Science and Technology of China, Hefei 230026, People's Republic of China}
\affiliation{Synergetic Innovation Center of Quantum Information and Quantum Physics, University of Science and Technology of China, Hefei 230026, People's Republic of China}

\date{\today}

\begin{abstract}
We report the experimental measurement of the winding number in an unitary chiral quantum walk. %Firstly, we propose the adoption of birefringent crystal for realizing the spin-orbit coupling in discrete time quantum walks.
Fundamentally, the spin-orbit coupling in discrete time quantum walks is implemented via birefringent crystal collinearly cut based on time-multiplexing scheme. Our protocol is compact and avoids extra loss, making it suitable for realizing genuine single-photon quantum walks at a large-scale. By adopting heralded single-photon as the walker and with a high time resolution technology in single-photon detection, we carry out a 50-step Hadamard discrete-time quantum walk with high fidelity up to 0.948$\bm{\pm}$0.007. Particularly, we can reconstruct the complete wave-function of the walker that starts the walk in a single lattice site through local tomography of each site. Through a Fourier transform, the wave-function in quasi-momentum space can be obtained. With this ability, we propose and report a method to reconstruct the eigenvectors of the system Hamiltonian in quasi-momentum space and directly read out the winding numbers in different topological phases (trivial and non-trivial) in the presence of chiral symmetry. By introducing nonequivalent time-frames, we show that the whole topological phases in periodically driven system can also be characterized by two different winding numbers. Our method can also be extended to the high winding number situation.
\end{abstract}

%\pacs{}
%\keywords{}
\maketitle
%\tableofcontents
%\section{Introduction}
\label{intro}
%\emph{Introduction.---}
Topological phases show distinctive characteristics that are beyond the Landau-Ginzburg-Wilson paradigm of symmetry breaking\,\cite{Landau1980,Sachdev2011,Wilson1974}. In topological phases, there is no local order parameter exhibited in the conventional phases due to symmetry breaking, and these phases are distinguished by some non-local topological invariants\,\cite{Sheng2006} of their ground state wave-functions which are quantized and robust. The classification of the topological phases is one of the main tasks in this field. Significant progresses have been made toward their complete classification in recent years\,\cite{Chen2012}. For a non-interaction system with some symmetries, such as particle-hole, time-reversal or chiral symmetry, a ``period table" of topological phases has been given\,\cite{Schnyder2008,Kitaev2009}. Understanding the topological phases not only is a fundamental problem but also has potential important applications in quantum information for their robustness. Such topological features have been explored in condensed matter systems\,\cite{Bednorz1986,Tsui1982,Laughlin1983}, high-energy physics\,\cite{Jackiw1976}, photonic systems\,\cite{Wang2009,Lu2016} and atomic physics\,\cite{Dauphin2013,Jotzu2014,Goldman2016}. Nevertheless, direct measurement of the topological invariants in these systems remains a difficult task.

Quantum walks (QWs)\,\cite{Aharonov1993}, which naturally couple the spin and movement of particles, provide a unique platform to investigate the topological phases in non-interaction systems with certain symmetries\,\cite{Kitagawa2010a}. It can reveal all topological phases occurring in one- and two-dimensional systems of non-interaction particles. Topologically protected bound states\,\cite{Kitagawa2012b} between two different bulk topological phases have been observed in QWs. %In addition, phase transitions between two different topological phases, which do not break the symmetry and are thus difficult to detect by the conventional method, have been observed in QWs by
The statistical moments of the final distribution in the lattice are introduced for monitoring phase transitions between trivial and non-trivial topological phases\,\cite{Cardano2016}. However, exploring topological phases by the topological invariants of the bulk observable is still a current challenge \cite{Tarasinski2014,Hauke2014,Flaschner2016} and only few examples in QWs have been demonstrated. Probing the topological invariants through the accumulated Berry phase in a Bloch oscillating type QW has been proposed in \cite{Ramasesh2017} and realized subsequently in \cite{Flurin2017}. The topology in the non-Hermitian QWs has also been investigated recently\,\cite{Rudner2009,Zeuner2015,Rakovszky2017,Xiao2017,Zhan2017}. It is proposed to understand the topology in QWs using scattering theory\,\cite{Tarasinski2014}, which has been demonstrated recently in \cite{Barkhofen2017}. As a periodically driven one-dimensional(1D) system, the complemented determination of its topology has been well studied theoretically\,\cite{Asboth2012,Asboth2013,Asboth2014,Obuse2015,Cedzich2016a,Cedzich2016b} with an experimental demonstration presented recently through measuring the mean chiral displacement (proportional to the Zak phase) in an unitary QW\,\cite{Cardano2017}.

\begin{figure}
\centering
\includegraphics[width=0.47\textwidth]{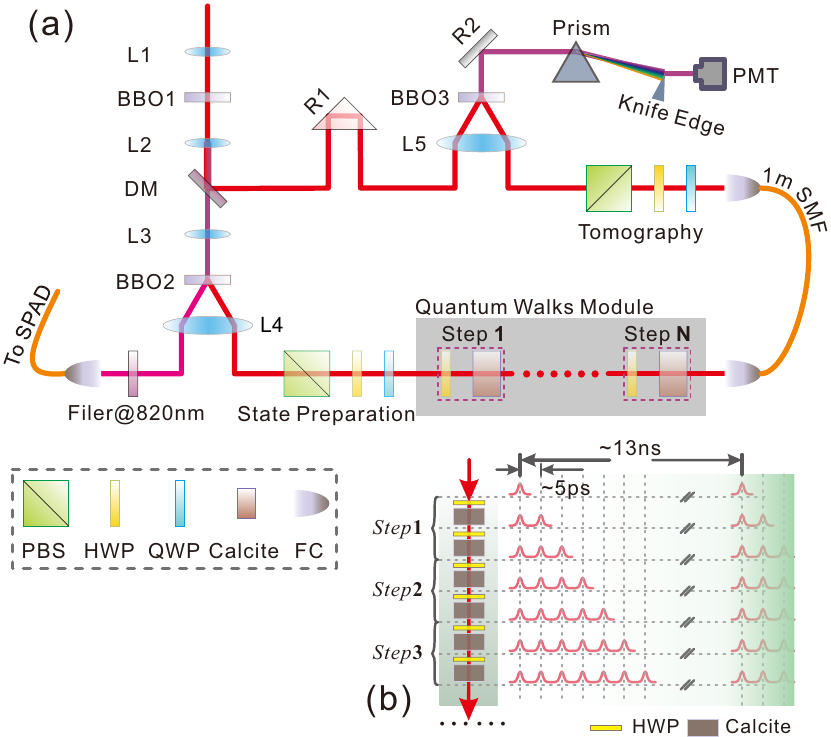}
%\begin{minipage}[b]{0.5\textwidth}
%\includegraphics[width=1\textwidth]{figures/Setup.pdf}
%\end{minipage}
%\begin{minipage}[b]{0.5\textwidth}
%\includegraphics[width=1\textwidth]{figures/Probability1.pdf} \
%\includegraphics[width=1\textwidth]{figures/Probability2.pdf}
%\end{minipage}
\caption{(a) diagram of the experimental setup with (b) presenting the compact time-multiplexing protocol by using birefringent crystals. The system contains four parts: 1. Second harmonic generation in BBO1 for obtaining the ultraviolet pulse; 2. Preparation of the heralded single photons by adopting the beam-like SPDC in BBO2; 3. Time-multiplexing QWs realized by birefringent crystals; 4. Ultra-fast detection of the arriving time of the single photons via up-conversion in BBO3. BBO: $\beta$-BaB$_2$O$_4$, L: lens, DM: dichroic mirror, PBS: polarization dependent beam splitter, HWP: half-wave plate, QWP: quarter-wave plate, R: reflector, FC: fibre collimator, SMF: single mode fibre, PMT: photomultiplier tubes, SPAD: single-photon avalanche diode. Detail descriptions are provided in supplementary.}
\label{Fig.Setup}
\end{figure}

In this work, we report a novel platform for QWs and experimentally show that it is a powerful tool to investigate the topological characters in 1D discrete time quantum walks (DTQWs), such as, directly obtaining the topological invariants. There have been many approaches to realize photonic QWs (reviewed in\,\cite{Wang2014}). Our scheme is based on the framework of time multiplexing, which is free of mode matching. Although there have been many reports\,\cite{Schreiber2010,Schreiber2011,Schreiber2012,Jeong2013}, one obstacle limiting their developments in scale and rejecting the employment of genuine single photons as the walker is the extra loss induced by the loop structure. For overcoming this obstacle, we propose using birefringent crystal to realize spin-orbit coupling. With its features of compactness and free of extra loss, our method can be used to realize large-scale QWs of single photons. The photon's polarization is adopted as the coin space and the coin tossing in each step can be varied arbitrarily by wave plates. Experimentally, for detecting and analyzing the single-photon signals with time intervals of a few picoseconds, we adopt the frequency up conversion single-photon detector for tomography of the spinor state in each lattice site\,\footnote{See Supplemental Material for brief description, which includes Refs.\,\cite{OConnor2012,Hadfield2009,Trebino2012,Ma2012}.}.

Particularly, the complete wave-function of the system %in which the walker is started from a single site
can be reconstructed in real space by local tomography of the spinor state in each site\,\cite{James2001} and interference measurements between the nearest neighbor sites\,\footnote{In Ref.\,\cite{Cardano2015}, reconstructing the full wave-function in usual interferometer based QWs was deemed to a challenge}, then the wave-function in quasi-momentum space can be obtained via Fourier transform. % and measuring the phase difference between nearest neighbor site.
Concretely, suppose the system after $t$-step walks is in state $|\Psi_t\rangle = \sum_{x}p_t(x)e^{-i\phi_t(x)}|\psi_t(x)\rangle\otimes|x\rangle$. For each site $x$ there is a normalized local spinor state $\ket{\psi_t(x)}$ with a complex amplitude $p_t(x)e^{-i\phi_t(x)}$ ($p_t(x)$ is a real valued quantity), where $\ket{\psi_t(x)} = \cos[\theta_t(x)/2]\ket{\uparrow} + e^{i\delta_t(x)}\sin[\theta_t(x)/2]\ket{\downarrow}$ with $\theta_t(x)\in[0,\pi]$ and $\delta_t(x)\in[0,2\pi)$. Experimentally, we obtain these parameters through three steps (details are given in supplementary): firstly, we perform local tomography on the spin for each site and get a count set $\mathcal{S}$; then we shift all of the spin-up components a step backward and perform local tomography again to obtain an additional count set $\tilde{\mathcal{S}}$ (local interference measurements); finally, we carry out a numerical global optimization program based on simulated annealing algorithm to find the optimal pure state $\ket{\Psi_t}$ from the two count sets $\mathcal{S}+\tilde{\mathcal{S}}$. Actually, our reconstruction method is an interferometric approach\,\cite{Barkhofen2017,Pears2017}. In addition, it can be systematically improved by increasing the rank of the target density matrices\,\cite {Gross2010,Zhao2017} (current pure state situation corresponds to rank 1).

\begin{figure*}
  \centering
  % Requires \usepackage{graphicx}
  %\includegraphics[width=7in]{figures/ExpResults50.pdf}\\
  \includegraphics[width=0.95\textwidth]{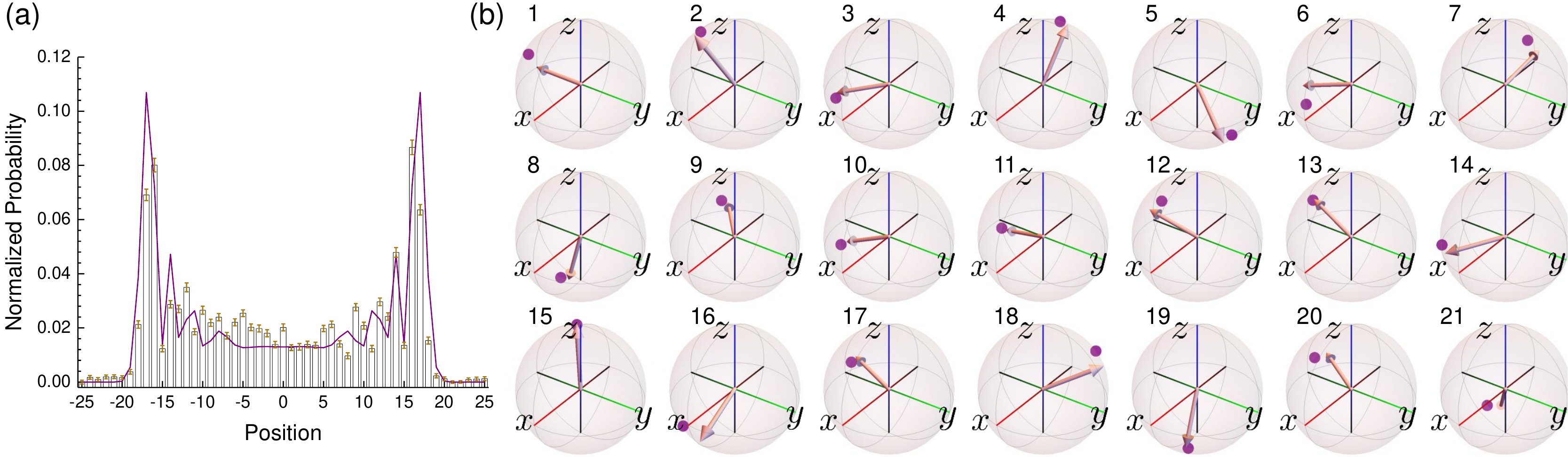}
  \caption{(a) histogram of the experimentally measured final probability distribution after a 50-step Hadamard QW starting from the origin with the spinor initialized in $\ket{L} = \frac{1}{\sqrt{2}}(\ket{H}+i\ket{V})$ (purple line guides the theoretical expectation). (b) shows the reconstructed spinor states (purple points) for each quasi-momentum $k$ (increasing in the first Brillouin zone indexed by the number) after a 20-step split-step QW with $\theta_1 = \pi/8,~\theta_2 = \pi/18$. The theoretical vectors are shown with the arrows.}\label{Fig.SpinState}
\end{figure*}

The layout of the apparatus is shown in Fig.\,\ref{Fig.Setup}. The signal photon of the photon-pair from beam-like spontaneous parametric down conversion (SPDC)\,\cite{Huang2011} is adopted as the walker, whose polarization can be initialized to any state by a typical polarizer. The birefringent crystals in collinear cut are used to realize the spin-orbit couplings and HWPs are inserted between them for coin tossing. The walker's final spinor states for all sites (time bins with equal interval 5\,$ps$) are analyzed with a polarizer and the corresponding amplitudes are measured with an up-conversion single-photon detector\,\cite{VanDevender2003}. Our method in realizing time-multiplexing QWs can avoid the extra loss in previous schemes and its collinear feature, as a result, without mode matching, can guarantee the visibility and stability in the interference. Here, we report, for the first time in a photonic system, a conventional Hadamard DTQW\,\cite{Aharonov1993} of heralded single photons on the scale of 50-step with high fidelity. The final probability distribution is presented in Fig.\,\ref{Fig.SpinState}(a). We compare the experimental distribution $P_\text{exp}$ with its theoretical prediction $P_\text{th}$ by the classical indicator similarity\,\cite{Schreiber2012}, defined as $S = [\sum_x\sqrt{P_\text{th}(x)P_\text{exp}(x)}]^2$, which gives 0.948$\bm{\pm}$0.007. The number of steps achieved here parallels that in other scalable QW platform based on ultracold atoms\,\cite{Andrea2014}.  In Fig.\,\ref{Fig.SpinState}(b), we further show the reconstructed spinor states for every quasi-momentum $k$ after a 20-step QW starting from the origin. The fidelity defined as $|\langle\Psi_\text{th}|\Psi_\text{exp}\rangle|^2$ is $0.960\pm0.005$. Here we only consider the statistical noise in estimating the errors of the fidelity and the final probability distribution in Fig.\,\ref{Fig.SpinState}. %Here all the errors are numerically estimated with only considering the statistical noise.

%With the ability of reconstructing the full wave-function in lattice, we can subsequently obtain the walker's spinor states in quasi-momentum space by performing Fourier transform onto the reconstructed wave-function in lattice.

%\emph{The distribution of the spinor states in the quasi-momentum space on the Bloch sphere gives some qualitative indicator of the topology phase as shown in the insets in Fig.\,\ref{Fig.ThePre}(c)\&(d).} Further, with several (at least 3) different spinor states in the quasi-momentum space which are obtained from different steps of a walker starting from the same single lattice, we can determine the effective $\bm{n}(k)$ which is the spinor eigenvectors for each momentum $k$. The winding number of the corresponding phase can then be directly read out from $\bm{n}(k)$.

\begin{figure}
\centering
\includegraphics[width=0.48\textwidth]{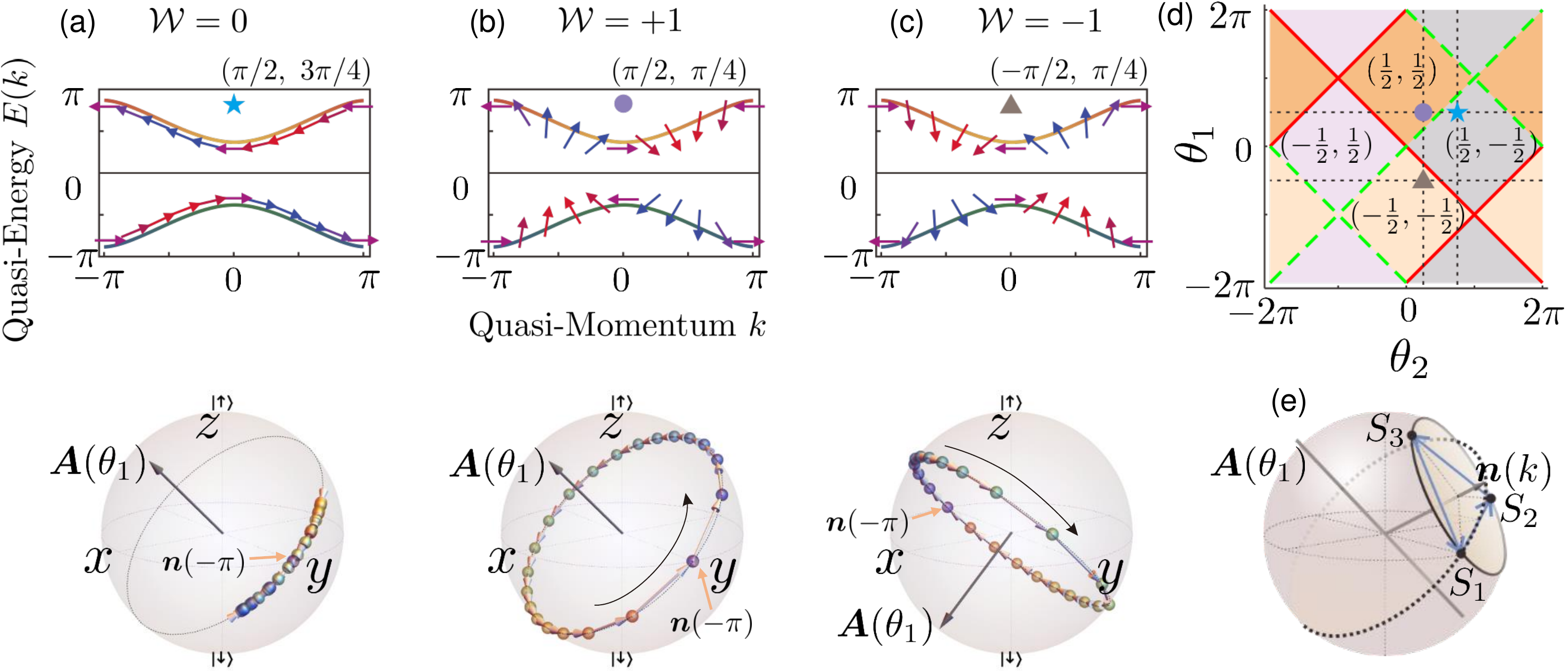}
%\subfigure[Phase Diagram]{\includegraphics[width=0.3\textwidth]{figures/PhaseDiagram.pdf}}
%%\mbox{\hspace{0.5cm}}
%\subfigure[Statistical Moment]{\includegraphics[width=0.3\textwidth]{figures/Moments.pdf}}
%\subfigure[Rotation Angle]{\includegraphics[width=0.3\textwidth]{figures/MaximalAngle.pdf}}
%\\
%\subfigure[Spinor eigenstates for $\theta_2=\pi/4$]{\includegraphics[width=0.29\textwidth]{figures/BlochVector1.pdf}}
%%\mbox{\hspace{0.5cm}}
%\subfigure[Final state in quasi-momentum space for $\theta_2=\pi/4$]{\includegraphics[width=0.3\textwidth]{figures/FinalState1.pdf}}
%\subfigure[Final distribution for $\theta_2=\pi/4$]{\includegraphics[width=0.3\textwidth]{figures/distribution1.pdf}}
%\\
%\subfigure[Spinor eigenstates for $\theta_2=3\pi/4$]{\includegraphics[width=0.29\textwidth]{figures/BlochVector2.pdf}}
%%\mbox{\hspace{0.5cm}}
%\subfigure[Final state in quasi-momentum space for $\theta_2=3\pi/4$]{\includegraphics[width=0.3\textwidth]{figures/FinalState2.pdf}}
%%\mbox{\hspace{0.5cm}}
%\subfigure[Final distribution for $\theta_2=3\pi/4$]{\includegraphics[width=0.3\textwidth]{figures/distribution2.pdf}}
%\renewcommand{\figurename}{Fig}
\caption{Topological characterizations of a split-step DTQW in the standard time-frame. (a)-(c) depict the energy bands (upper) and the corresponding eigenvectors (bottom) in three different topological phases, whose diagram formed by the various combinations of the two rotation angles is presented in (d). The guiding arrows in the energy band show the different winding feature with their exact forms $\bm n(k)$ presented in the bottom Bloch spheres. (e) sketches the method for reconstructing $\bm{n}(k)$ through final states from three different steps $(S_1$-$S_3)$.}
\label{Fig.ThePre}
\end{figure}

%\section*{Results}
%\subsection{Theoretical model for DTQWs.}
Generally, topological characters of a system can be determined by its topological protected edge modes or bulk invariants. There exists a correspondence between the edge modes and the bulk invariants. For translation invariant bulk systems, the winding number is known as a good invariant for classifying the topological phases\,\cite{Kitaev2009}. For the periodically driven system (QW here), the complete classification of the topological phases can be realized by two winding numbers corresponding to two QWs with different time-frames \,\cite{Asboth2012,Asboth2013,Asboth2014,Obuse2015,Cedzich2016a,Cedzich2016b}. Thus, determining the winding number plays the key role to understand the topological phases in QWs. %The direct way is to get all eigenvectors of the system in the same band. Fortunately, in our current platform, the wave-function in momentum space for each step can be reconstructed, which can be further used to determine the eigenvectors.
The direct way would be to measure all eigenvectors of the system for a given band. Fortunately, we do not need to populate a single band since we can obtain the same information by studying the evolution of the wave-function in momentum space for different steps. It needs to be noted that a large-scale QW is necessary to obtain enough eigenvectors to determine winding number.

To demonstrate the benefits of our current platform in investigating the topological phases, we adopt the split-step protocol\,\cite{Kitagawa2010a}, where the time evolution operator defined in the standard time-frame is $U(\theta_1,\theta_2) = T_-R(\theta_2)T_+R(\theta_1)$. $T_\pm  \coloneqq \sum_x(\ket{x\pm1}\bra{x}\otimes\ket{\pm}\bra{\pm} + \ket{x}\bra{x}\otimes\ket{\mp}\bra{\mp})$ is the shift operator with $\ket{+}=\ket{\uparrow}, \ket{-} = \ket{\downarrow}$ and $R(\theta_{1(2)})$ is the coin tossing (here we can take it as a spinor rotation along $\sigma_y$ without loss of generality). Completely classifying the topological phases in such a periodically driven system, we apply two nonequivalent shifted time-frames as \,\cite{Asboth2013,Obuse2015}: $U'(\theta_1,\theta_2) = R(\theta_1/2)T_-R(\theta_2)T_+R(\theta_1/2) ~\&~ U''(\theta_1,\theta_2) = R(\theta_2/2)T_+R(\theta_1)T_-R(\theta_2/2)$. In these two time-frames,  a chiral symmetry ($\sigma_x$), which is independent of the system's parameters, can then be defined. Therefore, we can define the corresponding winding numbers $\nu'$ and $\nu''$ through the Berry phases accumulated by the eigenvectors $\bm{n}(k)$ as the quasi-momentum $k$ runs from $-\pi$ to $\pi$ in the first Brillouin zone. Two invariants $\nu_0 = (\nu' + \nu'')/2~\& ~\nu_\pi = (\nu'-\nu'')/2$ can be subsequently introduced to completely determine the topology. Numerical results of the energy band and the corresponding eigenvectors (which are constrained in a plane defined by $A(\theta_1)$ in the standard time-frame) for different topological phases are shown in Fig.\,\ref{Fig.ThePre}.

As a periodically driven system, the time evolution operator $U$ can be represented by the evolution of an effective Hamiltonian as: $U(\theta_1, \theta_2)=e^{-iH_\text{eff}(\theta_1, \theta_2)}$ with $H_\text{eff}(\theta_1, \theta_2) = \int_{-\pi}^\pi dk[E(k)\bm{n}(k)\cdot\bm{\hat\sigma}]\otimes|k\rangle\langle k|.$ %(the detailed formula of $E(k)$ and $\bm{n}(k)$ for $U'$ and $U''$ can be found in supplementary).
Physically, $\bm{n}(k)$ defines an axis of rotation for each $k$, around which the spinor states are revolved, as shown in Fig.\,\ref{Fig.ThePre}(e). As a result, the spinor states for each given $k$ with different steps will be constrained to lie on a plane that is perpendicular to $\bm{n}(k)$ (the normal vector of the plane). Conversely, for fixed $\theta_{1(2)}$, the eigenvectors $\bm{n}(k)$ for each $k$ can be determined with the plane formed by at least three different spinor states (which can be reconstructed in our platform). The sign for $\bm{n}(k)$ (`plus' or `minus' corresponding to two eigenvectors) remains uncertain. Resorting to the continuation of $\bm{n}(k)$ in $k$ space and assuming the direction of $\bm{n}(k_0)$ (where $k_0$ can be arbitrarily chosen) is fixed, the entire eigenvectors can then be uniquely determined. With the obtained eigenvectors for every $k$, we can directly read out the winding number. Obviously, the method proposed here can be directly extended to the high winding number ($\geq 2$) situations. We need to note here that our method can be taken as a kind of dynamical measurement of system's eigenstates, which has been demonstrated in atoms system\,\cite{Flaschner2016}.

\begin{figure}
  \centering
  % Requires \usepackage{graphicx}
  %\includegraphics[width=7in]{figures/ExpResults50.pdf}\\
  %\includegraphics[width=0.5\textwidth]{figures/ExpResults50WithoutRepair.pdf}
  \includegraphics[width=0.48\textwidth]{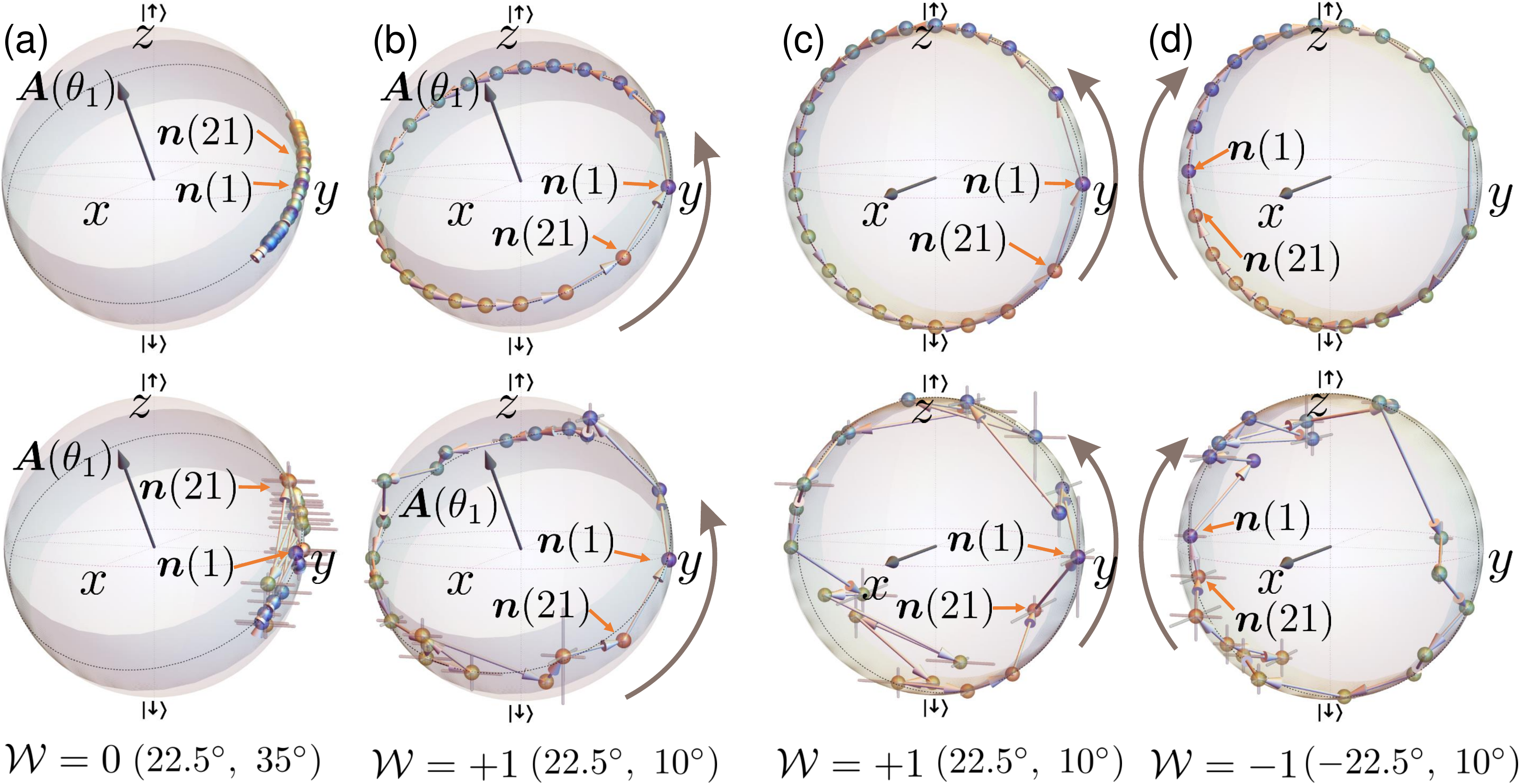}
  \caption{Theoretical predictions (upper rows) and experimental results (lower rows) for directly measuring the topological invariants with (a)\&(b) in a standard time-frame and (c)\&(d) in a shifted time-frame. The total size of the lattice is 21 after a 20-step QW.  Experimentally measured eigenvectors are given by the colored points on the Bloch sphere. The eigenvectors start winding from $\bm n(1)$ to $\bm n(21)$ with the direction presented by the arrow. In trivial phase (a), winding of $\bm n(k)$ forms an arc corresponding to a winding number $\mathcal{W} = 0$. In non-trivial phase (b)-(d), winding of $\bm n(k)$ forms an circle corresponding to a winding number $\mathcal{W} = \pm 1$, where the sign can be defined by the winding direction. The parameters $(\theta_1,~\theta_2)$ for each scenario are given at the bottom. Each experimentally measured $\bm n(k)$ is an average over ten different measurements and the translucent bars in each $\bm n(k)$ give the associated standard errors, which indicate the total noise.}\label{Fig.nk}
\end{figure}

In our experiment, we use three spinor states in the $k$-space for each $k$, i.e., the initial state (a single lattice site state), the relevant states reconstructed from the 1- and 20-step walks starting from the initial state, to form a plane and obtain its normal vector (the spinor eigenvector $\bm n(k)$). The resolution of the momentum space is determined by the largest steps--20. Concretely, we chose the parameters $(\theta_1,~\theta_2)$ to be $(22.5^\circ, 10^\circ)$ in non-trivial phase and $(22.5^\circ, 35^\circ)$ in trivial phase in the standard time-frame as examples. Resorting to the method introduced in the previous section, we obtained the entire $\bm n(k)$ and depicted these eigenvectors on the Bloch sphere in Fig.\,\ref{Fig.nk}(a)\&(b). Due to the imperfections in experiment, the reconstructed eigenvectors will extend to a certain range (depicted with the color band with $\pm\pi/10$ divergence) on the surface of the Bloch sphere instead of constrained on a plane (predicted in theory as shown in the upper row). However, in this practical situation, the winding of $\bm n(k)$ around the axis $\bm{A}(\theta_1)$ (in the standard time-frame) is also well defined. It can be calculated by projecting these eigenvectors to the plane which is perpendicular to $\bm{A}(\theta_1)$. Our experimental results clearly show the different winding features in the topological trivial and non-trivial phases.

To complete the classification of the non-trivial phases in the QWs, two winding numbers, $\nu'~\&~\nu''$, defined in two nonequivalent shifted time-frames\,\cite{Asboth2013,Obuse2015} are necessary. In the split-step protocol, the time evolution operators $U'$ and $U''$ are identical only by switching $\theta_1$ and $\theta_2$. In consequence, the critical requirement is to catch the feature that in non-trivial phases the winding number $\nu'$ is $+1$ for $\theta_1\in\{0,\pi\}$ and $-1$ for $\theta_1\in\{-\pi,0\}$ ($\nu''$ is trivial in this scenario)\,\cite{Obuse2015}. We further perform QWs in the shifted time-frame ($U'$) and reconstruct $\bm n(k)$ (shown in Fig.\ref{Fig.nk}(c)\&(d)). Our results show that the winding numbers of the parameters ($22.5^{\circ}$, $10^{\circ}$) and $(-22.5^{\circ}$, $10^{\circ}$) are different. With the winding numbers $\nu'$ and $\nu''$ obtained in the two nonequivalent time-frames, we can obtain the invariants $(\nu_{0}$, $\nu_\pi)$ and complete the classification of the topological phases in QWs.

One of the critical characters of topological phases is its robustness. We consider the winding of the eigenvectors $\bm{n}(k)$ in the presence of dynamic disorder, that is, the parameters $\theta_{1}$ randomly change its value at each step within a given range $\Delta\theta_{1}$. Without loss of generality, we consider the QW with a winding number $\mathcal{W} = -1$ in the shifted time-frame $U'$. The chiral symmetry is not destroyed in this dynamic disorder as its definition is independent of system parameters. We numerically investigate the influences of the disorder strength $\Delta\theta_{1} = 3^\circ,~4^\circ,~10^\circ$ and different steps of walks on the winding of eigenvectors (the results are presented in supplementary). Our simulations show that although the $\bm{n}(k)$ will diverge from the chiral plane defined by $\sigma_x$ and the divergence will increase with the increase of disorder strength and the steps of the walk, the robust of the winding number holds provided that the disorder strength is smaller than the gap size and the evolution time has not extended beyond the coherence time. Large disorder strength or long time evolution (beyond the coherent time) will significantly mix the system, which results in the failure of reading out the eigenvectors. Therefore, the robust of the winding number will be destroyed\,\cite{Obuse2011,Groh2016}.
In summary, we develop a photonic platform for realizing single-photon DTQWs with the ability of reconstructing the full final wave-function. Our scheme is robust in overcoming the visibility and stability problems experienced in previous trails. %, and thereby provides a new method for constructing quantum simulators based on large-scale QWs.
Additionally, based on the technology in extracting the full information of the system, we perform an experiment to obtain the system's spinor eigenvectors and directly read out the winding number of the bulk observable. The method proposed here is general and can be extended to determine high winding number topological phases. In prospect, our approach in directly measuring the topology based on reconstructing the spinor states in quasi-momentum space may be applicable to the investigation and observation of other classes of phase transitions in more complex quantum systems\,\cite{Heyl2013,Viyuela2014,Hu2016}, thus providing new perspectives for investigating the topology in physics.

%\begin{acknowledgments}
This work was supported by National Key Research and Development Program of China (Nos.\,2017YFA0304100, 2016YFA0302700), the National Natural Science Foundation of China (Nos.\,11474267, 61327901, 61322506, 11774335, 61725504), Key Research Program of Frontier Sciences, CAS (No.\,QYZDY-SSW-SLH003), the Fundamental Research Funds for the Central Universities (No.\,WK2470000026), the National Postdoctoral Program for Innovative Talents (No.\,BX201600146), China Postdoctoral Science Foundation (No.\,2017M612073) and Anhui Initiative in Quantum Information Technologies (No.\,AHY020100).
%\end{acknowledgments}

\bibliographystyle{apsrev4-1.bst}
\bibliography{references}% Produces the bibliography via BibTeX.

%merlin.mbs apsrev4-1.bst 2010-07-25 4.21a (PWD, AO, DPC) hacked
%Control: key (0)
%Control: author (72) initials jnrlst
%Control: editor formatted (1) identically to author
%Control: production of article title (-1) disabled
%Control: page (0) single
%Control: year (1) truncated
%Control: production of eprint (0) enabled
\providecommand{\noopsort}[1]{}\providecommand{\singleletter}[1]{#1}%
\begin{thebibliography}{62}%
\makeatletter
\providecommand \@ifxundefined [1]{%
 \@ifx{#1\undefined}
}%
\providecommand \@ifnum [1]{%
 \ifnum #1\expandafter \@firstoftwo
 \else \expandafter \@secondoftwo
 \fi
}%
\providecommand \@ifx [1]{%
 \ifx #1\expandafter \@firstoftwo
 \else \expandafter \@secondoftwo
 \fi
}%
\providecommand \natexlab [1]{#1}%
\providecommand \enquote  [1]{``#1''}%
\providecommand \bibnamefont  [1]{#1}%
\providecommand \bibfnamefont [1]{#1}%
\providecommand \citenamefont [1]{#1}%
\providecommand \href@noop [0]{\@secondoftwo}%
\providecommand \href [0]{\begingroup \@sanitize@url \@href}%
\providecommand \@href[1]{\@@startlink{#1}\@@href}%
\providecommand \@@href[1]{\endgroup#1\@@endlink}%
\providecommand \@sanitize@url [0]{\catcode `\\12\catcode `\$12\catcode
  `\&12\catcode `\#12\catcode `\^12\catcode `\_12\catcode `\%12\relax}%
\providecommand \@@startlink[1]{}%
\providecommand \@@endlink[0]{}%
\providecommand \url  [0]{\begingroup\@sanitize@url \@url }%
\providecommand \@url [1]{\endgroup\@href {#1}{\urlprefix }}%
\providecommand \urlprefix  [0]{URL }%
\providecommand \Eprint [0]{\href }%
\providecommand \doibase [0]{http://dx.doi.org/}%
\providecommand \selectlanguage [0]{\@gobble}%
\providecommand \bibinfo  [0]{\@secondoftwo}%
\providecommand \bibfield  [0]{\@secondoftwo}%
\providecommand \translation [1]{[#1]}%
\providecommand \BibitemOpen [0]{}%
\providecommand \bibitemStop [0]{}%
\providecommand \bibitemNoStop [0]{.\EOS\space}%
\providecommand \EOS [0]{\spacefactor3000\relax}%
\providecommand \BibitemShut  [1]{\csname bibitem#1\endcsname}%
\let\auto@bib@innerbib\@empty
%</preamble>
\bibitem [{\citenamefont {Landau}\ \emph {et~al.}(1980)\citenamefont {Landau},
  \citenamefont {Lifsh\texttoptiebar{its}},\ and\ \citenamefont
  {Pitaevski\u{\i}}}]{Landau1980}%
  \BibitemOpen
  \bibfield  {author} {\bibinfo {author} {\bibfnamefont {L.~D.}\ \bibnamefont
  {Landau}}, \bibinfo {author} {\bibfnamefont {E.~M.}\ \bibnamefont
  {Lifsh\texttoptiebar{its}}}, \ and\ \bibinfo {author} {\bibfnamefont {L.~P.}\
  \bibnamefont {Pitaevski\u{\i}}},\ }\href@noop {} {\emph {\bibinfo {title}
  {Statistical Physics}}},\ \bibinfo {edition} {3rd}\ ed.\ (\bibinfo
  {publisher} {Pergamon Press},\ \bibinfo {address} {New York},\ \bibinfo
  {year} {1980})\BibitemShut {NoStop}%
\bibitem [{\citenamefont {Sachdev}(2011)}]{Sachdev2011}%
  \BibitemOpen
  \bibfield  {author} {\bibinfo {author} {\bibfnamefont {S.}~\bibnamefont
  {Sachdev}},\ }\href@noop {} {\emph {\bibinfo {title} {Quantum phase
  transitions}}},\ \bibinfo {edition} {2nd}\ ed.\ (\bibinfo  {publisher}
  {Cambridge University Press},\ \bibinfo {address} {Cambridge, U.K.},\
  \bibinfo {year} {2011})\BibitemShut {NoStop}%
\bibitem [{\citenamefont {Wilson}\ and\ \citenamefont
  {Kogut}(1974)}]{Wilson1974}%
  \BibitemOpen
  \bibfield  {author} {\bibinfo {author} {\bibfnamefont {K.~G.}\ \bibnamefont
  {Wilson}}\ and\ \bibinfo {author} {\bibfnamefont {J.}~\bibnamefont {Kogut}},\
  }\href@noop {} {\bibfield  {journal} {\bibinfo  {journal} {Phys. Rep.}\
  }\textbf {\bibinfo {volume} {12}},\ \bibinfo {pages} {75} (\bibinfo {year}
  {1974})}\BibitemShut {NoStop}%
\bibitem [{\citenamefont {Sheng}\ \emph {et~al.}(2006)\citenamefont {Sheng},
  \citenamefont {Weng}, \citenamefont {Sheng},\ and\ \citenamefont
  {Haldane}}]{Sheng2006}%
  \BibitemOpen
  \bibfield  {author} {\bibinfo {author} {\bibfnamefont {D.~N.}\ \bibnamefont
  {Sheng}}, \bibinfo {author} {\bibfnamefont {Z.~Y.}\ \bibnamefont {Weng}},
  \bibinfo {author} {\bibfnamefont {L.}~\bibnamefont {Sheng}}, \ and\ \bibinfo
  {author} {\bibfnamefont {F.~D.~M.}\ \bibnamefont {Haldane}},\ }\href@noop {}
  {\bibfield  {journal} {\bibinfo  {journal} {Phys. Rev. Lett.}\ }\textbf
  {\bibinfo {volume} {97}},\ \bibinfo {pages} {036808} (\bibinfo {year}
  {2006})}\BibitemShut {NoStop}%
\bibitem [{\citenamefont {Chen}\ \emph {et~al.}(2012)\citenamefont {Chen},
  \citenamefont {Gu}, \citenamefont {Liu},\ and\ \citenamefont
  {Wen}}]{Chen2012}%
  \BibitemOpen
  \bibfield  {author} {\bibinfo {author} {\bibfnamefont {X.}~\bibnamefont
  {Chen}}, \bibinfo {author} {\bibfnamefont {Z.-C.}\ \bibnamefont {Gu}},
  \bibinfo {author} {\bibfnamefont {Z.-X.}\ \bibnamefont {Liu}}, \ and\
  \bibinfo {author} {\bibfnamefont {X.-G.}\ \bibnamefont {Wen}},\ }\href@noop
  {} {\bibfield  {journal} {\bibinfo  {journal} {Science}\ }\textbf {\bibinfo
  {volume} {338}},\ \bibinfo {pages} {1604} (\bibinfo {year}
  {2012})}\BibitemShut {NoStop}%
\bibitem [{\citenamefont {Schnyder}\ \emph {et~al.}(2008)\citenamefont
  {Schnyder}, \citenamefont {Ryu}, \citenamefont {Furusaki},\ and\
  \citenamefont {Ludwig}}]{Schnyder2008}%
  \BibitemOpen
  \bibfield  {author} {\bibinfo {author} {\bibfnamefont {A.~P.}\ \bibnamefont
  {Schnyder}}, \bibinfo {author} {\bibfnamefont {S.}~\bibnamefont {Ryu}},
  \bibinfo {author} {\bibfnamefont {A.}~\bibnamefont {Furusaki}}, \ and\
  \bibinfo {author} {\bibfnamefont {A.~W.~W.}\ \bibnamefont {Ludwig}},\
  }\href@noop {} {\bibfield  {journal} {\bibinfo  {journal} {Phys. Rev. B}\
  }\textbf {\bibinfo {volume} {78}},\ \bibinfo {pages} {195125} (\bibinfo
  {year} {2008})}\BibitemShut {NoStop}%
\bibitem [{\citenamefont {Kitaev}(2009)}]{Kitaev2009}%
  \BibitemOpen
  \bibfield  {author} {\bibinfo {author} {\bibfnamefont {A.}~\bibnamefont
  {Kitaev}},\ }\href@noop {} {\bibfield  {journal} {\bibinfo  {journal} {AIP
  Conf. Proc.}\ }\textbf {\bibinfo {volume} {1134}},\ \bibinfo {pages} {22}
  (\bibinfo {year} {2009})}\BibitemShut {NoStop}%
\bibitem [{\citenamefont {Bednorz}\ and\ \citenamefont
  {M\"uller}(1986)}]{Bednorz1986}%
  \BibitemOpen
  \bibfield  {author} {\bibinfo {author} {\bibfnamefont {J.~G.}\ \bibnamefont
  {Bednorz}}\ and\ \bibinfo {author} {\bibfnamefont {K.~A.}\ \bibnamefont
  {M\"uller}},\ }\href@noop {} {\bibfield  {journal} {\bibinfo  {journal}
  {Zeitschrift F\"ur Physik B - Condensed Matter}\ }\textbf {\bibinfo {volume}
  {64}},\ \bibinfo {pages} {189} (\bibinfo {year} {1986})}\BibitemShut
  {NoStop}%
\bibitem [{\citenamefont {Tsui}\ \emph {et~al.}(1982)\citenamefont {Tsui},
  \citenamefont {Stormer},\ and\ \citenamefont {Gossard}}]{Tsui1982}%
  \BibitemOpen
  \bibfield  {author} {\bibinfo {author} {\bibfnamefont {D.~C.}\ \bibnamefont
  {Tsui}}, \bibinfo {author} {\bibfnamefont {H.~L.}\ \bibnamefont {Stormer}}, \
  and\ \bibinfo {author} {\bibfnamefont {A.~C.}\ \bibnamefont {Gossard}},\
  }\href@noop {} {\bibfield  {journal} {\bibinfo  {journal} {Phys. Rev. Lett.}\
  }\textbf {\bibinfo {volume} {48}},\ \bibinfo {pages} {1559} (\bibinfo {year}
  {1982})}\BibitemShut {NoStop}%
\bibitem [{\citenamefont {Laughlin}(1983)}]{Laughlin1983}%
  \BibitemOpen
  \bibfield  {author} {\bibinfo {author} {\bibfnamefont {R.~B.}\ \bibnamefont
  {Laughlin}},\ }\href@noop {} {\bibfield  {journal} {\bibinfo  {journal}
  {Phys. Rev. Lett.}\ }\textbf {\bibinfo {volume} {50}},\ \bibinfo {pages}
  {1395} (\bibinfo {year} {1983})}\BibitemShut {NoStop}%
\bibitem [{\citenamefont {Jackiw}\ and\ \citenamefont
  {Rebbi}(1976)}]{Jackiw1976}%
  \BibitemOpen
  \bibfield  {author} {\bibinfo {author} {\bibfnamefont {R.}~\bibnamefont
  {Jackiw}}\ and\ \bibinfo {author} {\bibfnamefont {C.}~\bibnamefont {Rebbi}},\
  }\href@noop {} {\bibfield  {journal} {\bibinfo  {journal} {Phys. Rev. D}\
  }\textbf {\bibinfo {volume} {13}},\ \bibinfo {pages} {3398} (\bibinfo {year}
  {1976})}\BibitemShut {NoStop}%
\bibitem [{\citenamefont {Wang}\ \emph {et~al.}(2009)\citenamefont {Wang},
  \citenamefont {Chong}, \citenamefont {Joannopoulos},\ and\ \citenamefont
  {Solja\v{o}i\'{c}}}]{Wang2009}%
  \BibitemOpen
  \bibfield  {author} {\bibinfo {author} {\bibfnamefont {Z.}~\bibnamefont
  {Wang}}, \bibinfo {author} {\bibfnamefont {Y.}~\bibnamefont {Chong}},
  \bibinfo {author} {\bibfnamefont {J.~D.}\ \bibnamefont {Joannopoulos}}, \
  and\ \bibinfo {author} {\bibfnamefont {M.}~\bibnamefont {Solja\v{o}i\'{c}}},\
  }\href@noop {} {\bibfield  {journal} {\bibinfo  {journal} {Nature}\ }\textbf
  {\bibinfo {volume} {461}},\ \bibinfo {pages} {772} (\bibinfo {year}
  {2009})}\BibitemShut {NoStop}%
\bibitem [{\citenamefont {Lu}\ \emph {et~al.}(2016)\citenamefont {Lu},
  \citenamefont {Joannopoulos},\ and\ \citenamefont
  {Solja\v{o}i\'{c}}}]{Lu2016}%
  \BibitemOpen
  \bibfield  {author} {\bibinfo {author} {\bibfnamefont {L.}~\bibnamefont
  {Lu}}, \bibinfo {author} {\bibfnamefont {J.~D.}\ \bibnamefont
  {Joannopoulos}}, \ and\ \bibinfo {author} {\bibfnamefont {M.}~\bibnamefont
  {Solja\v{o}i\'{c}}},\ }\href@noop {} {\bibfield  {journal} {\bibinfo
  {journal} {Nat. Phys.}\ }\textbf {\bibinfo {volume} {12}},\ \bibinfo {pages}
  {626} (\bibinfo {year} {2016})}\BibitemShut {NoStop}%
\bibitem [{\citenamefont {Dauphin}\ and\ \citenamefont
  {Goldman}(2013)}]{Dauphin2013}%
  \BibitemOpen
  \bibfield  {author} {\bibinfo {author} {\bibfnamefont {A.}~\bibnamefont
  {Dauphin}}\ and\ \bibinfo {author} {\bibfnamefont {N.}~\bibnamefont
  {Goldman}},\ }\href@noop {} {\bibfield  {journal} {\bibinfo  {journal} {Phys.
  Rev. Lett.}\ }\textbf {\bibinfo {volume} {111}},\ \bibinfo {pages} {135302}
  (\bibinfo {year} {2013})}\BibitemShut {NoStop}%
\bibitem [{\citenamefont {Jotzu}\ \emph {et~al.}(2014)\citenamefont {Jotzu},
  \citenamefont {Messer}, \citenamefont {Desbuquois}, \citenamefont {Lebrat},
  \citenamefont {Uehlinger}, \citenamefont {Greif},\ and\ \citenamefont
  {Esslinger}}]{Jotzu2014}%
  \BibitemOpen
  \bibfield  {author} {\bibinfo {author} {\bibfnamefont {G.}~\bibnamefont
  {Jotzu}}, \bibinfo {author} {\bibfnamefont {M.}~\bibnamefont {Messer}},
  \bibinfo {author} {\bibfnamefont {R.}~\bibnamefont {Desbuquois}}, \bibinfo
  {author} {\bibfnamefont {M.}~\bibnamefont {Lebrat}}, \bibinfo {author}
  {\bibfnamefont {T.}~\bibnamefont {Uehlinger}}, \bibinfo {author}
  {\bibfnamefont {D.}~\bibnamefont {Greif}}, \ and\ \bibinfo {author}
  {\bibfnamefont {T.}~\bibnamefont {Esslinger}},\ }\href@noop {} {\bibfield
  {journal} {\bibinfo  {journal} {Nature}\ }\textbf {\bibinfo {volume} {515}},\
  \bibinfo {pages} {237} (\bibinfo {year} {2014})}\BibitemShut {NoStop}%
\bibitem [{\citenamefont {Goldman}\ \emph {et~al.}(2016)\citenamefont
  {Goldman}, \citenamefont {Budich},\ and\ \citenamefont
  {Zoller}}]{Goldman2016}%
  \BibitemOpen
  \bibfield  {author} {\bibinfo {author} {\bibfnamefont {N.}~\bibnamefont
  {Goldman}}, \bibinfo {author} {\bibfnamefont {J.~C.}\ \bibnamefont {Budich}},
  \ and\ \bibinfo {author} {\bibfnamefont {P.}~\bibnamefont {Zoller}},\
  }\href@noop {} {\bibfield  {journal} {\bibinfo  {journal} {Nat. Phys.}\
  }\textbf {\bibinfo {volume} {12}},\ \bibinfo {pages} {639} (\bibinfo {year}
  {2016})}\BibitemShut {NoStop}%
\bibitem [{\citenamefont {Aharonov}\ \emph {et~al.}(1993)\citenamefont
  {Aharonov}, \citenamefont {Davidovich},\ and\ \citenamefont
  {Zagury}}]{Aharonov1993}%
  \BibitemOpen
  \bibfield  {author} {\bibinfo {author} {\bibfnamefont {Y.}~\bibnamefont
  {Aharonov}}, \bibinfo {author} {\bibfnamefont {L.}~\bibnamefont
  {Davidovich}}, \ and\ \bibinfo {author} {\bibfnamefont {N.}~\bibnamefont
  {Zagury}},\ }\href@noop {} {\bibfield  {journal} {\bibinfo  {journal} {Phys.
  Rev. A}\ }\textbf {\bibinfo {volume} {48}},\ \bibinfo {pages} {1687}
  (\bibinfo {year} {1993})}\BibitemShut {NoStop}%
\bibitem [{\citenamefont {Kitagawa}\ \emph {et~al.}(2010)\citenamefont
  {Kitagawa}, \citenamefont {Rudner}, \citenamefont {Berg},\ and\ \citenamefont
  {Demler}}]{Kitagawa2010a}%
  \BibitemOpen
  \bibfield  {author} {\bibinfo {author} {\bibfnamefont {T.}~\bibnamefont
  {Kitagawa}}, \bibinfo {author} {\bibfnamefont {M.~S.}\ \bibnamefont
  {Rudner}}, \bibinfo {author} {\bibfnamefont {E.}~\bibnamefont {Berg}}, \ and\
  \bibinfo {author} {\bibfnamefont {E.}~\bibnamefont {Demler}},\ }\href@noop {}
  {\bibfield  {journal} {\bibinfo  {journal} {Phys. Rev. A}\ }\textbf {\bibinfo
  {volume} {82}},\ \bibinfo {pages} {033429} (\bibinfo {year}
  {2010})}\BibitemShut {NoStop}%
\bibitem [{\citenamefont {Kitagawa}\ \emph {et~al.}(2012)\citenamefont
  {Kitagawa}, \citenamefont {Broome}, \citenamefont {Fedrizzi}, \citenamefont
  {Rudner}, \citenamefont {Berg}, \citenamefont {Kassal}, \citenamefont
  {Aspuru-Guzik}, \citenamefont {Demler},\ and\ \citenamefont
  {White}}]{Kitagawa2012b}%
  \BibitemOpen
  \bibfield  {author} {\bibinfo {author} {\bibfnamefont {T.}~\bibnamefont
  {Kitagawa}}, \bibinfo {author} {\bibfnamefont {M.~A.}\ \bibnamefont
  {Broome}}, \bibinfo {author} {\bibfnamefont {A.}~\bibnamefont {Fedrizzi}},
  \bibinfo {author} {\bibfnamefont {M.~S.}\ \bibnamefont {Rudner}}, \bibinfo
  {author} {\bibfnamefont {E.}~\bibnamefont {Berg}}, \bibinfo {author}
  {\bibfnamefont {I.}~\bibnamefont {Kassal}}, \bibinfo {author} {\bibfnamefont
  {A.}~\bibnamefont {Aspuru-Guzik}}, \bibinfo {author} {\bibfnamefont
  {E.}~\bibnamefont {Demler}}, \ and\ \bibinfo {author} {\bibfnamefont {A.~G.}\
  \bibnamefont {White}},\ }\href@noop {} {\bibfield  {journal} {\bibinfo
  {journal} {Nat. Commun.}\ }\textbf {\bibinfo {volume} {3}},\ \bibinfo {pages}
  {882} (\bibinfo {year} {2012})}\BibitemShut {NoStop}%
\bibitem [{\citenamefont {Cardano}\ \emph {et~al.}(2016)\citenamefont
  {Cardano}, \citenamefont {Maffei}, \citenamefont {Massa}, \citenamefont
  {Piccirillo}, \citenamefont {de~Lisio}, \citenamefont {De~Filippis},
  \citenamefont {Cataudella}, \citenamefont {Santamato},\ and\ \citenamefont
  {Marrucci}}]{Cardano2016}%
  \BibitemOpen
  \bibfield  {author} {\bibinfo {author} {\bibfnamefont {F.}~\bibnamefont
  {Cardano}}, \bibinfo {author} {\bibfnamefont {M.}~\bibnamefont {Maffei}},
  \bibinfo {author} {\bibfnamefont {F.}~\bibnamefont {Massa}}, \bibinfo
  {author} {\bibfnamefont {B.}~\bibnamefont {Piccirillo}}, \bibinfo {author}
  {\bibfnamefont {C.}~\bibnamefont {de~Lisio}}, \bibinfo {author}
  {\bibfnamefont {G.}~\bibnamefont {De~Filippis}}, \bibinfo {author}
  {\bibfnamefont {V.}~\bibnamefont {Cataudella}}, \bibinfo {author}
  {\bibfnamefont {E.}~\bibnamefont {Santamato}}, \ and\ \bibinfo {author}
  {\bibfnamefont {L.}~\bibnamefont {Marrucci}},\ }\href@noop {} {\bibfield
  {journal} {\bibinfo  {journal} {Nat. Commun.}\ }\textbf {\bibinfo {volume}
  {7}},\ \bibinfo {pages} {11439} (\bibinfo {year} {2016})}\BibitemShut
  {NoStop}%
\bibitem [{\citenamefont {Tarasinski}\ \emph {et~al.}(2014)\citenamefont
  {Tarasinski}, \citenamefont {Asb\'oth},\ and\ \citenamefont
  {Dahlhaus}}]{Tarasinski2014}%
  \BibitemOpen
  \bibfield  {author} {\bibinfo {author} {\bibfnamefont {B.}~\bibnamefont
  {Tarasinski}}, \bibinfo {author} {\bibfnamefont {J.~K.}\ \bibnamefont
  {Asb\'oth}}, \ and\ \bibinfo {author} {\bibfnamefont {J.~P.}\ \bibnamefont
  {Dahlhaus}},\ }\href@noop {} {\bibfield  {journal} {\bibinfo  {journal}
  {Phys. Rev. A}\ }\textbf {\bibinfo {volume} {89}},\ \bibinfo {pages} {042327}
  (\bibinfo {year} {2014})}\BibitemShut {NoStop}%
\bibitem [{\citenamefont {Hauke}\ \emph {et~al.}(2014)\citenamefont {Hauke},
  \citenamefont {Lewenstein},\ and\ \citenamefont {Eckardt}}]{Hauke2014}%
  \BibitemOpen
  \bibfield  {author} {\bibinfo {author} {\bibfnamefont {P.}~\bibnamefont
  {Hauke}}, \bibinfo {author} {\bibfnamefont {M.}~\bibnamefont {Lewenstein}}, \
  and\ \bibinfo {author} {\bibfnamefont {A.}~\bibnamefont {Eckardt}},\
  }\href@noop {} {\bibfield  {journal} {\bibinfo  {journal} {Phys. Rev. Lett.}\
  }\textbf {\bibinfo {volume} {113}},\ \bibinfo {pages} {045303} (\bibinfo
  {year} {2014})}\BibitemShut {NoStop}%
\bibitem [{\citenamefont {Fl{\"a}schner}\ \emph {et~al.}(2016)\citenamefont
  {Fl{\"a}schner}, \citenamefont {Rem}, \citenamefont {Tarnowski},
  \citenamefont {Vogel}, \citenamefont {L{\"u}hmann}, \citenamefont
  {Sengstock},\ and\ \citenamefont {Weitenberg}}]{Flaschner2016}%
  \BibitemOpen
  \bibfield  {author} {\bibinfo {author} {\bibfnamefont {N.}~\bibnamefont
  {Fl{\"a}schner}}, \bibinfo {author} {\bibfnamefont {B.~S.}\ \bibnamefont
  {Rem}}, \bibinfo {author} {\bibfnamefont {M.}~\bibnamefont {Tarnowski}},
  \bibinfo {author} {\bibfnamefont {D.}~\bibnamefont {Vogel}}, \bibinfo
  {author} {\bibfnamefont {D.-S.}\ \bibnamefont {L{\"u}hmann}}, \bibinfo
  {author} {\bibfnamefont {K.}~\bibnamefont {Sengstock}}, \ and\ \bibinfo
  {author} {\bibfnamefont {C.}~\bibnamefont {Weitenberg}},\ }\href@noop {}
  {\bibfield  {journal} {\bibinfo  {journal} {Science}\ }\textbf {\bibinfo
  {volume} {352}},\ \bibinfo {pages} {1091} (\bibinfo {year}
  {2016})}\BibitemShut {NoStop}%
\bibitem [{\citenamefont {Ramasesh}\ \emph {et~al.}(2017)\citenamefont
  {Ramasesh}, \citenamefont {Flurin}, \citenamefont {Rudner}, \citenamefont
  {Siddiqi},\ and\ \citenamefont {Yao}}]{Ramasesh2017}%
  \BibitemOpen
  \bibfield  {author} {\bibinfo {author} {\bibfnamefont {V.~V.}\ \bibnamefont
  {Ramasesh}}, \bibinfo {author} {\bibfnamefont {E.}~\bibnamefont {Flurin}},
  \bibinfo {author} {\bibfnamefont {M.}~\bibnamefont {Rudner}}, \bibinfo
  {author} {\bibfnamefont {I.}~\bibnamefont {Siddiqi}}, \ and\ \bibinfo
  {author} {\bibfnamefont {N.~Y.}\ \bibnamefont {Yao}},\ }\href@noop {}
  {\bibfield  {journal} {\bibinfo  {journal} {Phys. Rev. Lett.}\ }\textbf
  {\bibinfo {volume} {118}},\ \bibinfo {pages} {130501} (\bibinfo {year}
  {2017})}\BibitemShut {NoStop}%
\bibitem [{\citenamefont {Flurin}\ \emph {et~al.}(2017)\citenamefont {Flurin},
  \citenamefont {Ramasesh}, \citenamefont {Hacohen-Gourgy}, \citenamefont
  {Martin}, \citenamefont {Yao},\ and\ \citenamefont {Siddiqi}}]{Flurin2017}%
  \BibitemOpen
  \bibfield  {author} {\bibinfo {author} {\bibfnamefont {E.}~\bibnamefont
  {Flurin}}, \bibinfo {author} {\bibfnamefont {V.~â.}\ \bibnamefont
  {Ramasesh}}, \bibinfo {author} {\bibfnamefont {S.}~\bibnamefont
  {Hacohen-Gourgy}}, \bibinfo {author} {\bibfnamefont {L.~â.}\ \bibnamefont
  {Martin}}, \bibinfo {author} {\bibfnamefont {N.~â.}\ \bibnamefont {Yao}}, \
  and\ \bibinfo {author} {\bibfnamefont {I.}~\bibnamefont {Siddiqi}},\
  }\href@noop {} {\bibfield  {journal} {\bibinfo  {journal} {Phys. Rev. X}\
  }\textbf {\bibinfo {volume} {7}},\ \bibinfo {pages} {031023} (\bibinfo {year}
  {2017})}\BibitemShut {NoStop}%
\bibitem [{\citenamefont {Rudner}\ and\ \citenamefont
  {Levitov}(2009)}]{Rudner2009}%
  \BibitemOpen
  \bibfield  {author} {\bibinfo {author} {\bibfnamefont {M.~S.}\ \bibnamefont
  {Rudner}}\ and\ \bibinfo {author} {\bibfnamefont {L.~S.}\ \bibnamefont
  {Levitov}},\ }\href@noop {} {\bibfield  {journal} {\bibinfo  {journal} {Phys.
  Rev. Lett.}\ }\textbf {\bibinfo {volume} {102}},\ \bibinfo {pages} {065703}
  (\bibinfo {year} {2009})}\BibitemShut {NoStop}%
\bibitem [{\citenamefont {Zeuner}\ \emph {et~al.}(2015)\citenamefont {Zeuner},
  \citenamefont {Rechtsman}, \citenamefont {Plotnik}, \citenamefont {Lumer},
  \citenamefont {Nolte}, \citenamefont {Rudner}, \citenamefont {Segev},\ and\
  \citenamefont {Szameit}}]{Zeuner2015}%
  \BibitemOpen
  \bibfield  {author} {\bibinfo {author} {\bibfnamefont {J.~M.}\ \bibnamefont
  {Zeuner}}, \bibinfo {author} {\bibfnamefont {M.~C.}\ \bibnamefont
  {Rechtsman}}, \bibinfo {author} {\bibfnamefont {Y.}~\bibnamefont {Plotnik}},
  \bibinfo {author} {\bibfnamefont {Y.}~\bibnamefont {Lumer}}, \bibinfo
  {author} {\bibfnamefont {S.}~\bibnamefont {Nolte}}, \bibinfo {author}
  {\bibfnamefont {M.~S.}\ \bibnamefont {Rudner}}, \bibinfo {author}
  {\bibfnamefont {M.}~\bibnamefont {Segev}}, \ and\ \bibinfo {author}
  {\bibfnamefont {A.}~\bibnamefont {Szameit}},\ }\href@noop {} {\bibfield
  {journal} {\bibinfo  {journal} {Phys. Rev. Lett.}\ }\textbf {\bibinfo
  {volume} {115}},\ \bibinfo {pages} {040402} (\bibinfo {year}
  {2015})}\BibitemShut {NoStop}%
\bibitem [{\citenamefont {Rakovszky}\ \emph {et~al.}(2017)\citenamefont
  {Rakovszky}, \citenamefont {Asb\'oth},\ and\ \citenamefont
  {Alberti}}]{Rakovszky2017}%
  \BibitemOpen
  \bibfield  {author} {\bibinfo {author} {\bibfnamefont {T.}~\bibnamefont
  {Rakovszky}}, \bibinfo {author} {\bibfnamefont {J.~K.}\ \bibnamefont
  {Asb\'oth}}, \ and\ \bibinfo {author} {\bibfnamefont {A.}~\bibnamefont
  {Alberti}},\ }\href@noop {} {\bibfield  {journal} {\bibinfo  {journal} {Phys.
  Rev. B}\ }\textbf {\bibinfo {volume} {95}},\ \bibinfo {pages} {201407}
  (\bibinfo {year} {2017})}\BibitemShut {NoStop}%
\bibitem [{\citenamefont {Xiao}\ \emph {et~al.}(2017)\citenamefont {Xiao},
  \citenamefont {Zhan}, \citenamefont {Bian}, \citenamefont {Wang},
  \citenamefont {Zhang}, \citenamefont {Wang}, \citenamefont {Li},
  \citenamefont {Mochizuki}, \citenamefont {Kim}, \citenamefont {Kawakami},
  \citenamefont {Yi}, \citenamefont {Obuse}, \citenamefont {Sanders},\ and\
  \citenamefont {Xue}}]{Xiao2017}%
  \BibitemOpen
  \bibfield  {author} {\bibinfo {author} {\bibfnamefont {L.}~\bibnamefont
  {Xiao}}, \bibinfo {author} {\bibfnamefont {X.}~\bibnamefont {Zhan}}, \bibinfo
  {author} {\bibfnamefont {Z.~H.}\ \bibnamefont {Bian}}, \bibinfo {author}
  {\bibfnamefont {K.~K.}\ \bibnamefont {Wang}}, \bibinfo {author}
  {\bibfnamefont {X.}~\bibnamefont {Zhang}}, \bibinfo {author} {\bibfnamefont
  {X.~P.}\ \bibnamefont {Wang}}, \bibinfo {author} {\bibfnamefont
  {J.}~\bibnamefont {Li}}, \bibinfo {author} {\bibfnamefont {K.}~\bibnamefont
  {Mochizuki}}, \bibinfo {author} {\bibfnamefont {D.}~\bibnamefont {Kim}},
  \bibinfo {author} {\bibfnamefont {N.}~\bibnamefont {Kawakami}}, \bibinfo
  {author} {\bibfnamefont {W.}~\bibnamefont {Yi}}, \bibinfo {author}
  {\bibfnamefont {H.}~\bibnamefont {Obuse}}, \bibinfo {author} {\bibfnamefont
  {B.~C.}\ \bibnamefont {Sanders}}, \ and\ \bibinfo {author} {\bibfnamefont
  {P.}~\bibnamefont {Xue}},\ }\href@noop {} {\bibfield  {journal} {\bibinfo
  {journal} {Nat. Phys.}\ }\textbf {\bibinfo {volume} {13}},\ \bibinfo {pages}
  {1117} (\bibinfo {year} {2017})}\BibitemShut {NoStop}%
\bibitem [{\citenamefont {Zhan}\ \emph {et~al.}(2017)\citenamefont {Zhan},
  \citenamefont {Xiao}, \citenamefont {Bian}, \citenamefont {Wang},
  \citenamefont {Qiu}, \citenamefont {Sanders}, \citenamefont {Yi},\ and\
  \citenamefont {Xue}}]{Zhan2017}%
  \BibitemOpen
  \bibfield  {author} {\bibinfo {author} {\bibfnamefont {X.}~\bibnamefont
  {Zhan}}, \bibinfo {author} {\bibfnamefont {L.}~\bibnamefont {Xiao}}, \bibinfo
  {author} {\bibfnamefont {Z.}~\bibnamefont {Bian}}, \bibinfo {author}
  {\bibfnamefont {K.}~\bibnamefont {Wang}}, \bibinfo {author} {\bibfnamefont
  {X.}~\bibnamefont {Qiu}}, \bibinfo {author} {\bibfnamefont {B.~C.}\
  \bibnamefont {Sanders}}, \bibinfo {author} {\bibfnamefont {W.}~\bibnamefont
  {Yi}}, \ and\ \bibinfo {author} {\bibfnamefont {P.}~\bibnamefont {Xue}},\
  }\href@noop {} {\bibfield  {journal} {\bibinfo  {journal} {Phys. Rev. Lett.}\
  }\textbf {\bibinfo {volume} {119}},\ \bibinfo {pages} {130501} (\bibinfo
  {year} {2017})}\BibitemShut {NoStop}%
\bibitem [{\citenamefont {Barkhofen}\ \emph {et~al.}(2017)\citenamefont
  {Barkhofen}, \citenamefont {Nitsche}, \citenamefont {Elster}, \citenamefont
  {Lorz}, \citenamefont {G\'abris}, \citenamefont {Jex},\ and\ \citenamefont
  {Silberhorn}}]{Barkhofen2017}%
  \BibitemOpen
  \bibfield  {author} {\bibinfo {author} {\bibfnamefont {S.}~\bibnamefont
  {Barkhofen}}, \bibinfo {author} {\bibfnamefont {T.}~\bibnamefont {Nitsche}},
  \bibinfo {author} {\bibfnamefont {F.}~\bibnamefont {Elster}}, \bibinfo
  {author} {\bibfnamefont {L.}~\bibnamefont {Lorz}}, \bibinfo {author}
  {\bibfnamefont {A.}~\bibnamefont {G\'abris}}, \bibinfo {author}
  {\bibfnamefont {I.}~\bibnamefont {Jex}}, \ and\ \bibinfo {author}
  {\bibfnamefont {C.}~\bibnamefont {Silberhorn}},\ }\href@noop {} {\bibfield
  {journal} {\bibinfo  {journal} {Phys. Rev. A}\ }\textbf {\bibinfo {volume}
  {96}},\ \bibinfo {pages} {033846} (\bibinfo {year} {2017})}\BibitemShut
  {NoStop}%
\bibitem [{\citenamefont {Asb\'{o}th}(2012)}]{Asboth2012}%
  \BibitemOpen
  \bibfield  {author} {\bibinfo {author} {\bibfnamefont {J.~K.}\ \bibnamefont
  {Asb\'{o}th}},\ }\href@noop {} {\bibfield  {journal} {\bibinfo  {journal}
  {Phys. Rev. B}\ }\textbf {\bibinfo {volume} {86}},\ \bibinfo {pages} {195414}
  (\bibinfo {year} {2012})}\BibitemShut {NoStop}%
\bibitem [{\citenamefont {Asb\'{o}th}\ and\ \citenamefont
  {Obuse}(2013)}]{Asboth2013}%
  \BibitemOpen
  \bibfield  {author} {\bibinfo {author} {\bibfnamefont {J.~K.}\ \bibnamefont
  {Asb\'{o}th}}\ and\ \bibinfo {author} {\bibfnamefont {H.}~\bibnamefont
  {Obuse}},\ }\href@noop {} {\bibfield  {journal} {\bibinfo  {journal} {Phys.
  Rev. B}\ }\textbf {\bibinfo {volume} {88}},\ \bibinfo {pages} {121406}
  (\bibinfo {year} {2013})}\BibitemShut {NoStop}%
\bibitem [{\citenamefont {Asb\'{o}th}\ \emph {et~al.}(2014)\citenamefont
  {Asb\'{o}th}, \citenamefont {Tarasinski},\ and\ \citenamefont
  {Delplace}}]{Asboth2014}%
  \BibitemOpen
  \bibfield  {author} {\bibinfo {author} {\bibfnamefont {J.~K.}\ \bibnamefont
  {Asb\'{o}th}}, \bibinfo {author} {\bibfnamefont {B.}~\bibnamefont
  {Tarasinski}}, \ and\ \bibinfo {author} {\bibfnamefont {P.}~\bibnamefont
  {Delplace}},\ }\href@noop {} {\bibfield  {journal} {\bibinfo  {journal}
  {Phys. Rev. B}\ }\textbf {\bibinfo {volume} {90}},\ \bibinfo {pages} {125143}
  (\bibinfo {year} {2014})}\BibitemShut {NoStop}%
\bibitem [{\citenamefont {Obuse}\ \emph {et~al.}(2015)\citenamefont {Obuse},
  \citenamefont {Asb\'{o}th}, \citenamefont {Nishimura},\ and\ \citenamefont
  {Kawakami}}]{Obuse2015}%
  \BibitemOpen
  \bibfield  {author} {\bibinfo {author} {\bibfnamefont {H.}~\bibnamefont
  {Obuse}}, \bibinfo {author} {\bibfnamefont {J.~K.}\ \bibnamefont
  {Asb\'{o}th}}, \bibinfo {author} {\bibfnamefont {Y.}~\bibnamefont
  {Nishimura}}, \ and\ \bibinfo {author} {\bibfnamefont {N.}~\bibnamefont
  {Kawakami}},\ }\href@noop {} {\bibfield  {journal} {\bibinfo  {journal}
  {Phys. Rev. B}\ }\textbf {\bibinfo {volume} {92}},\ \bibinfo {pages} {045424}
  (\bibinfo {year} {2015})}\BibitemShut {NoStop}%
\bibitem [{\citenamefont {Cedzich}\ \emph
  {et~al.}(2016{\natexlab{a}})\citenamefont {Cedzich}, \citenamefont {Geib},
  \citenamefont {Gr\"unbaum}, \citenamefont {Stahl}, \citenamefont
  {Vel\'azquez}, \citenamefont {Werner},\ and\ \citenamefont
  {Werner}}]{Cedzich2016a}%
  \BibitemOpen
  \bibfield  {author} {\bibinfo {author} {\bibfnamefont {C.}~\bibnamefont
  {Cedzich}}, \bibinfo {author} {\bibfnamefont {T.}~\bibnamefont {Geib}},
  \bibinfo {author} {\bibfnamefont {F.~A.}\ \bibnamefont {Gr\"unbaum}},
  \bibinfo {author} {\bibfnamefont {C.}~\bibnamefont {Stahl}}, \bibinfo
  {author} {\bibfnamefont {L.}~\bibnamefont {Vel\'azquez}}, \bibinfo {author}
  {\bibfnamefont {A.~H.}\ \bibnamefont {Werner}}, \ and\ \bibinfo {author}
  {\bibfnamefont {R.~F.}\ \bibnamefont {Werner}},\ }\href@noop {} {\bibfield
  {journal} {\bibinfo  {journal} {arXiv:1611.04439}\ } (\bibinfo {year}
  {2016}{\natexlab{a}})}\BibitemShut {NoStop}%
\bibitem [{\citenamefont {Cedzich}\ \emph
  {et~al.}(2016{\natexlab{b}})\citenamefont {Cedzich}, \citenamefont
  {Gr\"unbaum}, \citenamefont {Stahl}, \citenamefont {Vel\'azquez},
  \citenamefont {Werner},\ and\ \citenamefont {Werner}}]{Cedzich2016b}%
  \BibitemOpen
  \bibfield  {author} {\bibinfo {author} {\bibfnamefont {C.}~\bibnamefont
  {Cedzich}}, \bibinfo {author} {\bibfnamefont {F.~A.}\ \bibnamefont
  {Gr\"unbaum}}, \bibinfo {author} {\bibfnamefont {C.}~\bibnamefont {Stahl}},
  \bibinfo {author} {\bibfnamefont {L.}~\bibnamefont {Vel\'azquez}}, \bibinfo
  {author} {\bibfnamefont {A.~H.}\ \bibnamefont {Werner}}, \ and\ \bibinfo
  {author} {\bibfnamefont {R.~F.}\ \bibnamefont {Werner}},\ }\href@noop {}
  {\bibfield  {journal} {\bibinfo  {journal} {J. Phys. A - Math. Theor.}\
  }\textbf {\bibinfo {volume} {49}},\ \bibinfo {pages} {21LT01} (\bibinfo
  {year} {2016}{\natexlab{b}})}\BibitemShut {NoStop}%
\bibitem [{\citenamefont {Cardano}\ \emph {et~al.}(2017)\citenamefont
  {Cardano}, \citenamefont {D'Errico}, \citenamefont {Dauphin}, \citenamefont
  {Maffei}, \citenamefont {Piccirillo}, \citenamefont {de~Lisio}, \citenamefont
  {De~Filippis}, \citenamefont {Cataudella}, \citenamefont {Santamato},
  \citenamefont {Marrucci}, \citenamefont {Lewenstein},\ and\ \citenamefont
  {Massignan}}]{Cardano2017}%
  \BibitemOpen
  \bibfield  {author} {\bibinfo {author} {\bibfnamefont {F.}~\bibnamefont
  {Cardano}}, \bibinfo {author} {\bibfnamefont {A.}~\bibnamefont {D'Errico}},
  \bibinfo {author} {\bibfnamefont {A.}~\bibnamefont {Dauphin}}, \bibinfo
  {author} {\bibfnamefont {M.}~\bibnamefont {Maffei}}, \bibinfo {author}
  {\bibfnamefont {B.}~\bibnamefont {Piccirillo}}, \bibinfo {author}
  {\bibfnamefont {C.}~\bibnamefont {de~Lisio}}, \bibinfo {author}
  {\bibfnamefont {G.}~\bibnamefont {De~Filippis}}, \bibinfo {author}
  {\bibfnamefont {V.}~\bibnamefont {Cataudella}}, \bibinfo {author}
  {\bibfnamefont {E.}~\bibnamefont {Santamato}}, \bibinfo {author}
  {\bibfnamefont {L.}~\bibnamefont {Marrucci}}, \bibinfo {author}
  {\bibfnamefont {M.}~\bibnamefont {Lewenstein}}, \ and\ \bibinfo {author}
  {\bibfnamefont {P.}~\bibnamefont {Massignan}},\ }\href@noop {} {\bibfield
  {journal} {\bibinfo  {journal} {Nat. Commun.}\ }\textbf {\bibinfo {volume}
  {8}},\ \bibinfo {pages} {15516} (\bibinfo {year} {2017})}\BibitemShut
  {NoStop}%
\bibitem [{\citenamefont {Wang}\ and\ \citenamefont
  {Manouchehri}(2014)}]{Wang2014}%
  \BibitemOpen
  \bibfield  {author} {\bibinfo {author} {\bibfnamefont {J.}~\bibnamefont
  {Wang}}\ and\ \bibinfo {author} {\bibfnamefont {K.}~\bibnamefont
  {Manouchehri}},\ }\href@noop {} {\emph {\bibinfo {title} {Physical
  Implementation of Quantum Walks}}}\ (\bibinfo  {publisher} {Springer-Verlag,
  New York},\ \bibinfo {year} {2014})\BibitemShut {NoStop}%
\bibitem [{\citenamefont {Schreiber}\ \emph {et~al.}(2010)\citenamefont
  {Schreiber}, \citenamefont {Cassemiro}, \citenamefont {Poto\v{c}ek},
  \citenamefont {G\'abris}, \citenamefont {Mosley}, \citenamefont {Andersson},
  \citenamefont {Jex},\ and\ \citenamefont {Silberhorn}}]{Schreiber2010}%
  \BibitemOpen
  \bibfield  {author} {\bibinfo {author} {\bibfnamefont {A.}~\bibnamefont
  {Schreiber}}, \bibinfo {author} {\bibfnamefont {K.~N.}\ \bibnamefont
  {Cassemiro}}, \bibinfo {author} {\bibfnamefont {V.}~\bibnamefont
  {Poto\v{c}ek}}, \bibinfo {author} {\bibfnamefont {A.}~\bibnamefont
  {G\'abris}}, \bibinfo {author} {\bibfnamefont {P.~J.}\ \bibnamefont
  {Mosley}}, \bibinfo {author} {\bibfnamefont {E.}~\bibnamefont {Andersson}},
  \bibinfo {author} {\bibfnamefont {I.}~\bibnamefont {Jex}}, \ and\ \bibinfo
  {author} {\bibfnamefont {C.}~\bibnamefont {Silberhorn}},\ }\href@noop {}
  {\bibfield  {journal} {\bibinfo  {journal} {Phys. Rev. Lett.}\ }\textbf
  {\bibinfo {volume} {104}},\ \bibinfo {pages} {050502} (\bibinfo {year}
  {2010})}\BibitemShut {NoStop}%
\bibitem [{\citenamefont {Schreiber}\ \emph {et~al.}(2011)\citenamefont
  {Schreiber}, \citenamefont {Cassemiro}, \citenamefont {Poto\v{c}ek},
  \citenamefont {G\'abris}, \citenamefont {Jex},\ and\ \citenamefont
  {Silberhorn}}]{Schreiber2011}%
  \BibitemOpen
  \bibfield  {author} {\bibinfo {author} {\bibfnamefont {A.}~\bibnamefont
  {Schreiber}}, \bibinfo {author} {\bibfnamefont {K.~N.}\ \bibnamefont
  {Cassemiro}}, \bibinfo {author} {\bibfnamefont {V.}~\bibnamefont
  {Poto\v{c}ek}}, \bibinfo {author} {\bibfnamefont {A.}~\bibnamefont
  {G\'abris}}, \bibinfo {author} {\bibfnamefont {I.}~\bibnamefont {Jex}}, \
  and\ \bibinfo {author} {\bibfnamefont {C.}~\bibnamefont {Silberhorn}},\
  }\href@noop {} {\bibfield  {journal} {\bibinfo  {journal} {Phys. Rev. Lett.}\
  }\textbf {\bibinfo {volume} {106}},\ \bibinfo {pages} {180403} (\bibinfo
  {year} {2011})}\BibitemShut {NoStop}%
\bibitem [{\citenamefont {Schreiber}\ \emph {et~al.}(2012)\citenamefont
  {Schreiber}, \citenamefont {G\'abris}, \citenamefont {Rohde}, \citenamefont
  {Laiho}, \citenamefont {\v{S}tefa\v{n}\'ak}, \citenamefont {Poto\v{c}ek},
  \citenamefont {Hamilton}, \citenamefont {Jex},\ and\ \citenamefont
  {Silberhorn}}]{Schreiber2012}%
  \BibitemOpen
  \bibfield  {author} {\bibinfo {author} {\bibfnamefont {A.}~\bibnamefont
  {Schreiber}}, \bibinfo {author} {\bibfnamefont {A.}~\bibnamefont {G\'abris}},
  \bibinfo {author} {\bibfnamefont {P.~P.}\ \bibnamefont {Rohde}}, \bibinfo
  {author} {\bibfnamefont {K.}~\bibnamefont {Laiho}}, \bibinfo {author}
  {\bibfnamefont {M.}~\bibnamefont {\v{S}tefa\v{n}\'ak}}, \bibinfo {author}
  {\bibfnamefont {V.}~\bibnamefont {Poto\v{c}ek}}, \bibinfo {author}
  {\bibfnamefont {C.}~\bibnamefont {Hamilton}}, \bibinfo {author}
  {\bibfnamefont {I.}~\bibnamefont {Jex}}, \ and\ \bibinfo {author}
  {\bibfnamefont {C.}~\bibnamefont {Silberhorn}},\ }\href@noop {} {\bibfield
  {journal} {\bibinfo  {journal} {Science}\ }\textbf {\bibinfo {volume}
  {336}},\ \bibinfo {pages} {55} (\bibinfo {year} {2012})}\BibitemShut
  {NoStop}%
\bibitem [{\citenamefont {Jeong}\ \emph {et~al.}(2013)\citenamefont {Jeong},
  \citenamefont {Di~Franco}, \citenamefont {Lim}, \citenamefont {Kim},\ and\
  \citenamefont {Kim}}]{Jeong2013}%
  \BibitemOpen
  \bibfield  {author} {\bibinfo {author} {\bibfnamefont {Y.~C.}\ \bibnamefont
  {Jeong}}, \bibinfo {author} {\bibfnamefont {C.}~\bibnamefont {Di~Franco}},
  \bibinfo {author} {\bibfnamefont {H.~T.}\ \bibnamefont {Lim}}, \bibinfo
  {author} {\bibfnamefont {M.~S.}\ \bibnamefont {Kim}}, \ and\ \bibinfo
  {author} {\bibfnamefont {Y.~H.}\ \bibnamefont {Kim}},\ }\href@noop {}
  {\bibfield  {journal} {\bibinfo  {journal} {Nat. Commun.}\ }\textbf {\bibinfo
  {volume} {4}},\ \bibinfo {pages} {2471} (\bibinfo {year} {2013})}\BibitemShut
  {NoStop}%
\bibitem [{Note1()}]{Note1}%
  \BibitemOpen
  \bibinfo {note} {See Supplemental Material for brief description, which
  includes Refs.\protect \tmspace +\thinmuskip {.1667em}\cite
  {OConnor2012,Hadfield2009,Trebino2012,Ma2012}.}\BibitemShut {Stop}%
\bibitem [{\citenamefont {James}\ \emph {et~al.}(2001)\citenamefont {James},
  \citenamefont {Kwiat}, \citenamefont {Munro},\ and\ \citenamefont
  {White}}]{James2001}%
  \BibitemOpen
  \bibfield  {author} {\bibinfo {author} {\bibfnamefont {D.~F.~V.}\
  \bibnamefont {James}}, \bibinfo {author} {\bibfnamefont {P.~G.}\ \bibnamefont
  {Kwiat}}, \bibinfo {author} {\bibfnamefont {W.~J.}\ \bibnamefont {Munro}}, \
  and\ \bibinfo {author} {\bibfnamefont {A.~G.}\ \bibnamefont {White}},\ }\href
  {\doibase 10.1103/PhysRevA.64.052312} {\bibfield  {journal} {\bibinfo
  {journal} {Phys. Rev. A}\ }\textbf {\bibinfo {volume} {64}},\ \bibinfo
  {pages} {052312} (\bibinfo {year} {2001})}\BibitemShut {NoStop}%
\bibitem [{Note2()}]{Note2}%
  \BibitemOpen
  \bibinfo {note} {In Ref.\protect \tmspace +\thinmuskip {.1667em}\cite
  {Cardano2015}, reconstructing the full wave-function in usual interferometer
  based QWs was deemed to a challenge}\BibitemShut {NoStop}%
\bibitem [{\citenamefont {Pears~Stefano}\ \emph {et~al.}(2017)\citenamefont
  {Pears~Stefano}, \citenamefont {Reb\'on}, \citenamefont {Ledesma},\ and\
  \citenamefont {Iemmi}}]{Pears2017}%
  \BibitemOpen
  \bibfield  {author} {\bibinfo {author} {\bibfnamefont {Q.}~\bibnamefont
  {Pears~Stefano}}, \bibinfo {author} {\bibfnamefont {L.}~\bibnamefont
  {Reb\'on}}, \bibinfo {author} {\bibfnamefont {S.}~\bibnamefont {Ledesma}}, \
  and\ \bibinfo {author} {\bibfnamefont {C.}~\bibnamefont {Iemmi}},\ }\href
  {https://link.aps.org/doi/10.1103/PhysRevA.96.062328} {\bibfield  {journal}
  {\bibinfo  {journal} {Phys. Rev. A}\ }\textbf {\bibinfo {volume} {96}},\
  \bibinfo {pages} {062328} (\bibinfo {year} {2017})}\BibitemShut {NoStop}%
\bibitem [{\citenamefont {Gross}\ \emph {et~al.}(2010)\citenamefont {Gross},
  \citenamefont {Liu}, \citenamefont {Flammia}, \citenamefont {Becker},\ and\
  \citenamefont {Eisert}}]{Gross2010}%
  \BibitemOpen
  \bibfield  {author} {\bibinfo {author} {\bibfnamefont {D.}~\bibnamefont
  {Gross}}, \bibinfo {author} {\bibfnamefont {Y.-K.}\ \bibnamefont {Liu}},
  \bibinfo {author} {\bibfnamefont {S.~T.}\ \bibnamefont {Flammia}}, \bibinfo
  {author} {\bibfnamefont {S.}~\bibnamefont {Becker}}, \ and\ \bibinfo {author}
  {\bibfnamefont {J.}~\bibnamefont {Eisert}},\ }\href@noop {} {\bibfield
  {journal} {\bibinfo  {journal} {Phys. Rev. Lett.}\ }\textbf {\bibinfo
  {volume} {105}},\ \bibinfo {pages} {150401} (\bibinfo {year}
  {2010})}\BibitemShut {NoStop}%
\bibitem [{\citenamefont {Zhao}\ \emph {et~al.}(2017)\citenamefont {Zhao},
  \citenamefont {Hou}, \citenamefont {Xiang}, \citenamefont {Han},
  \citenamefont {Li},\ and\ \citenamefont {Guo}}]{Zhao2017}%
  \BibitemOpen
  \bibfield  {author} {\bibinfo {author} {\bibfnamefont {Y.-Y.}\ \bibnamefont
  {Zhao}}, \bibinfo {author} {\bibfnamefont {Z.}~\bibnamefont {Hou}}, \bibinfo
  {author} {\bibfnamefont {G.-Y.}\ \bibnamefont {Xiang}}, \bibinfo {author}
  {\bibfnamefont {Y.-J.}\ \bibnamefont {Han}}, \bibinfo {author} {\bibfnamefont
  {C.-F.}\ \bibnamefont {Li}}, \ and\ \bibinfo {author} {\bibfnamefont {G.-C.}\
  \bibnamefont {Guo}},\ }\href@noop {} {\bibfield  {journal} {\bibinfo
  {journal} {Opt. Express}\ }\textbf {\bibinfo {volume} {25}},\ \bibinfo
  {pages} {9010} (\bibinfo {year} {2017})}\BibitemShut {NoStop}%
\bibitem [{\citenamefont {Huang}\ \emph {et~al.}(2011)\citenamefont {Huang},
  \citenamefont {Liu}, \citenamefont {Peng}, \citenamefont {Li}, \citenamefont
  {Li}, \citenamefont {Li},\ and\ \citenamefont {Guo}}]{Huang2011}%
  \BibitemOpen
  \bibfield  {author} {\bibinfo {author} {\bibfnamefont {Y.~F.}\ \bibnamefont
  {Huang}}, \bibinfo {author} {\bibfnamefont {B.~H.}\ \bibnamefont {Liu}},
  \bibinfo {author} {\bibfnamefont {L.}~\bibnamefont {Peng}}, \bibinfo {author}
  {\bibfnamefont {Y.~H.}\ \bibnamefont {Li}}, \bibinfo {author} {\bibfnamefont
  {L.}~\bibnamefont {Li}}, \bibinfo {author} {\bibfnamefont {C.~F.}\
  \bibnamefont {Li}}, \ and\ \bibinfo {author} {\bibfnamefont {G.~C.}\
  \bibnamefont {Guo}},\ }\href@noop {} {\bibfield  {journal} {\bibinfo
  {journal} {Nat. Commun.}\ }\textbf {\bibinfo {volume} {2}},\ \bibinfo {pages}
  {546} (\bibinfo {year} {2011})}\BibitemShut {NoStop}%
\bibitem [{\citenamefont {VanDevender}\ and\ \citenamefont
  {Kwiat}(2003)}]{VanDevender2003}%
  \BibitemOpen
  \bibfield  {author} {\bibinfo {author} {\bibfnamefont {A.~P.}\ \bibnamefont
  {VanDevender}}\ and\ \bibinfo {author} {\bibfnamefont {P.~G.}\ \bibnamefont
  {Kwiat}},\ }in\ \href@noop {} {\emph {\bibinfo {booktitle} {AeroSense
  2003}}},\ Vol.\ \bibinfo {volume} {5105}\ (\bibinfo  {publisher} {SPIE},\
  \bibinfo {year} {2003})\ p.~\bibinfo {pages} {9}\BibitemShut {NoStop}%
\bibitem [{\citenamefont {Andrea}\ \emph {et~al.}(2014)\citenamefont {Andrea},
  \citenamefont {Wolfgang}, \citenamefont {Reinhard},\ and\ \citenamefont
  {Dieter}}]{Andrea2014}%
  \BibitemOpen
  \bibfield  {author} {\bibinfo {author} {\bibfnamefont {A.}~\bibnamefont
  {Andrea}}, \bibinfo {author} {\bibfnamefont {A.}~\bibnamefont {Wolfgang}},
  \bibinfo {author} {\bibfnamefont {W.}~\bibnamefont {Reinhard}}, \ and\
  \bibinfo {author} {\bibfnamefont {M.}~\bibnamefont {Dieter}},\ }\href@noop {}
  {\bibfield  {journal} {\bibinfo  {journal} {New J. Phys.}\ }\textbf {\bibinfo
  {volume} {16}},\ \bibinfo {pages} {123052} (\bibinfo {year}
  {2014})}\BibitemShut {NoStop}%
\bibitem [{\citenamefont {Obuse}\ and\ \citenamefont
  {Kawakami}(2011)}]{Obuse2011}%
  \BibitemOpen
  \bibfield  {author} {\bibinfo {author} {\bibfnamefont {H.}~\bibnamefont
  {Obuse}}\ and\ \bibinfo {author} {\bibfnamefont {N.}~\bibnamefont
  {Kawakami}},\ }\href {https://link.aps.org/doi/10.1103/PhysRevB.84.195139}
  {\bibfield  {journal} {\bibinfo  {journal} {Phys. Rev. B}\ }\textbf {\bibinfo
  {volume} {84}},\ \bibinfo {pages} {195139} (\bibinfo {year}
  {2011})}\BibitemShut {NoStop}%
\bibitem [{\citenamefont {Groh}\ \emph {et~al.}(2016)\citenamefont {Groh},
  \citenamefont {Brakhane}, \citenamefont {Alt}, \citenamefont {Meschede},
  \citenamefont {Asb\'oth},\ and\ \citenamefont {Alberti}}]{Groh2016}%
  \BibitemOpen
  \bibfield  {author} {\bibinfo {author} {\bibfnamefont {T.}~\bibnamefont
  {Groh}}, \bibinfo {author} {\bibfnamefont {S.}~\bibnamefont {Brakhane}},
  \bibinfo {author} {\bibfnamefont {W.}~\bibnamefont {Alt}}, \bibinfo {author}
  {\bibfnamefont {D.}~\bibnamefont {Meschede}}, \bibinfo {author}
  {\bibfnamefont {J.~K.}\ \bibnamefont {Asb\'oth}}, \ and\ \bibinfo {author}
  {\bibfnamefont {A.}~\bibnamefont {Alberti}},\ }\href {\doibase
  10.1103/PhysRevA.94.013620} {\bibfield  {journal} {\bibinfo  {journal} {Phys.
  Rev. A}\ }\textbf {\bibinfo {volume} {94}},\ \bibinfo {pages} {013620}
  (\bibinfo {year} {2016})}\BibitemShut {NoStop}%
\bibitem [{\citenamefont {Heyl}\ \emph {et~al.}(2013)\citenamefont {Heyl},
  \citenamefont {Polkovnikov},\ and\ \citenamefont {Kehrein}}]{Heyl2013}%
  \BibitemOpen
  \bibfield  {author} {\bibinfo {author} {\bibfnamefont {M.}~\bibnamefont
  {Heyl}}, \bibinfo {author} {\bibfnamefont {A.}~\bibnamefont {Polkovnikov}}, \
  and\ \bibinfo {author} {\bibfnamefont {S.}~\bibnamefont {Kehrein}},\ }\href
  {https://link.aps.org/doi/10.1103/PhysRevLett.110.135704} {\bibfield
  {journal} {\bibinfo  {journal} {Phys. Rev. Lett.}\ }\textbf {\bibinfo
  {volume} {110}},\ \bibinfo {pages} {135704} (\bibinfo {year}
  {2013})}\BibitemShut {NoStop}%
\bibitem [{\citenamefont {Viyuela}\ \emph {et~al.}(2014)\citenamefont
  {Viyuela}, \citenamefont {Rivas},\ and\ \citenamefont
  {Martin-Delgado}}]{Viyuela2014}%
  \BibitemOpen
  \bibfield  {author} {\bibinfo {author} {\bibfnamefont {O.}~\bibnamefont
  {Viyuela}}, \bibinfo {author} {\bibfnamefont {A.}~\bibnamefont {Rivas}}, \
  and\ \bibinfo {author} {\bibfnamefont {M.~A.}\ \bibnamefont
  {Martin-Delgado}},\ }\href@noop {} {\bibfield  {journal} {\bibinfo  {journal}
  {Phys. Rev. Lett.}\ }\textbf {\bibinfo {volume} {112}},\ \bibinfo {pages}
  {130401} (\bibinfo {year} {2014})}\BibitemShut {NoStop}%
\bibitem [{\citenamefont {Hu}\ \emph {et~al.}(2016)\citenamefont {Hu},
  \citenamefont {Zoller},\ and\ \citenamefont {Budich}}]{Hu2016}%
  \BibitemOpen
  \bibfield  {author} {\bibinfo {author} {\bibfnamefont {Y.}~\bibnamefont
  {Hu}}, \bibinfo {author} {\bibfnamefont {P.}~\bibnamefont {Zoller}}, \ and\
  \bibinfo {author} {\bibfnamefont {J.~C.}\ \bibnamefont {Budich}},\
  }\href@noop {} {\bibfield  {journal} {\bibinfo  {journal} {Phys. Rev. Lett.}\
  }\textbf {\bibinfo {volume} {117}},\ \bibinfo {pages} {126803} (\bibinfo
  {year} {2016})}\BibitemShut {NoStop}%
\bibitem [{\citenamefont {O'Connor}(2012)}]{OConnor2012}%
  \BibitemOpen
  \bibfield  {author} {\bibinfo {author} {\bibfnamefont {D.}~\bibnamefont
  {O'Connor}},\ }\href@noop {} {\emph {\bibinfo {title} {Time-correlated single
  photon counting}}}\ (\bibinfo  {publisher} {Academic Press},\ \bibinfo {year}
  {2012})\BibitemShut {NoStop}%
\bibitem [{\citenamefont {Hadfield}(2009)}]{Hadfield2009}%
  \BibitemOpen
  \bibfield  {author} {\bibinfo {author} {\bibfnamefont {R.~H.}\ \bibnamefont
  {Hadfield}},\ }\href@noop {} {\bibfield  {journal} {\bibinfo  {journal} {Nat.
  Photonics}\ }\textbf {\bibinfo {volume} {3}},\ \bibinfo {pages} {696}
  (\bibinfo {year} {2009})}\BibitemShut {NoStop}%
\bibitem [{\citenamefont {Trebino}(2012)}]{Trebino2012}%
  \BibitemOpen
  \bibfield  {author} {\bibinfo {author} {\bibfnamefont {R.}~\bibnamefont
  {Trebino}},\ }\href@noop {} {\emph {\bibinfo {title} {Frequency-resolved
  optical gating: the measurement of ultrashort laser pulses}}}\ (\bibinfo
  {publisher} {Springer Science \& Business Media},\ \bibinfo {year}
  {2012})\BibitemShut {NoStop}%
\bibitem [{\citenamefont {Ma}\ \emph {et~al.}(2012)\citenamefont {Ma},
  \citenamefont {Slattery},\ and\ \citenamefont {Tang}}]{Ma2012}%
  \BibitemOpen
  \bibfield  {author} {\bibinfo {author} {\bibfnamefont {L.~J.}\ \bibnamefont
  {Ma}}, \bibinfo {author} {\bibfnamefont {O.}~\bibnamefont {Slattery}}, \ and\
  \bibinfo {author} {\bibfnamefont {X.}~\bibnamefont {Tang}},\ }\href@noop {}
  {\bibfield  {journal} {\bibinfo  {journal} {Physics Reports-Review Section of
  Physics Letters}\ }\textbf {\bibinfo {volume} {521}},\ \bibinfo {pages} {69}
  (\bibinfo {year} {2012})}\BibitemShut {NoStop}%
\bibitem [{\citenamefont {Cardano}\ \emph {et~al.}(2015)\citenamefont
  {Cardano}, \citenamefont {Massa}, \citenamefont {Qassim}, \citenamefont
  {Karimi}, \citenamefont {Slussarenko}, \citenamefont {Paparo}, \citenamefont
  {de~Lisio}, \citenamefont {Sciarrino}, \citenamefont {Santamato},
  \citenamefont {Boyd},\ and\ \citenamefont {Marrucci}}]{Cardano2015}%
  \BibitemOpen
  \bibfield  {author} {\bibinfo {author} {\bibfnamefont {F.}~\bibnamefont
  {Cardano}}, \bibinfo {author} {\bibfnamefont {F.}~\bibnamefont {Massa}},
  \bibinfo {author} {\bibfnamefont {H.}~\bibnamefont {Qassim}}, \bibinfo
  {author} {\bibfnamefont {E.}~\bibnamefont {Karimi}}, \bibinfo {author}
  {\bibfnamefont {S.}~\bibnamefont {Slussarenko}}, \bibinfo {author}
  {\bibfnamefont {D.}~\bibnamefont {Paparo}}, \bibinfo {author} {\bibfnamefont
  {C.}~\bibnamefont {de~Lisio}}, \bibinfo {author} {\bibfnamefont
  {F.}~\bibnamefont {Sciarrino}}, \bibinfo {author} {\bibfnamefont
  {E.}~\bibnamefont {Santamato}}, \bibinfo {author} {\bibfnamefont {R.~W.}\
  \bibnamefont {Boyd}}, \ and\ \bibinfo {author} {\bibfnamefont
  {L.}~\bibnamefont {Marrucci}},\ }\href@noop {} {\bibfield  {journal}
  {\bibinfo  {journal} {Sci. Adv.}\ }\textbf {\bibinfo {volume} {1}},\ \bibinfo
  {pages} {e1500087} (\bibinfo {year} {2015})}\BibitemShut {NoStop}%
\end{thebibliography}%

\newpage
\appendix
\section{Supplementary for \emph{Measuring the Winding Number in a Large-Scale Chiral Quantum Walk}}

\section{Experimental time multiplexing photonic DTQWs}
\subsection{DTQWs in time domain}
In this work, the implementation of DTQW is based on the time multiplexing protocol\,\cite{Wang2014}. However, for overcoming the problem of the extra loss, birefringent crystals are used to implement the spin-orbit coupling instead of the asymmetric Mach-Zehnder interferometers. Heralded single photon generated from SPDC is employed as the walker. The polarization degree of the photon is employed as the coin space, such that its coin state can be optionally rotated via wave plates. The arriving time of the photon, encoded in time bin, acts as the position space. One step QW is realized by a module composed of a half wave plate (HWP) and one piece of birefringent crystal. The coin rotation operator can be written as
\begin{equation}
\hat R_\text{HWP}(\theta) = e^{-i2\theta\hat\sigma_y}\hat\sigma_z,
\end{equation}
where $\theta$ is the rotation angle of the optical axis of the HWP and $\hat\sigma_i~(i\in\{x,y,z\})$ denote the Pauli matrices. The eigenstates of the coin are $|H\rangle$ and $|V\rangle$, corresponding to the horizontal and vertical polarization respectively, with the condition $\hat\sigma_z|H\rangle = |H\rangle$ and $\hat\sigma_z|V\rangle = -|V\rangle$. The birefringence causes the horizontal components to travel faster inside the crystal than the vertical one. As a result, after passing through the crystal the photon in state $\ket{H}$ moves a step forward. Considering the dispersion after passing through a large number of crystals and the fact that the time bin encoding the position of the walker in reality is a single pulse with a typical duration of a few hundred femtoseconds, such a shift in time should be sufficiently large to distinguish the neighborhood pulses at last. The magnitude of the polarization-dependent time shift by the birefringent crystal depends on the crystal length and the cut angle. In our experiment, for introducing as weak dispersion as possible with sufficiently large birefringence, calcite crystal is adopted for its high birefringence index (0.167 at 800\,\emph{nm}). The length is chosen to be 8.98\,\emph{mm} with its optical axis parallel to incident plane, such that the time shift is designed to be 5\,\emph{ps} for one-step.

\subsection{Heralded single photon adopted as the walker}
The time multiplexing protocol requires pulse photons, which can be obtained by attenuating a pulse laser or modulating a continuous laser with an optical chopper. Considering the tradeoff between the operation on the time bins and the final analysis in time domain, the pulse duration covers a range from tens to thousands of picoseconds, reaching even a few microseconds. In our experiment, for adopting a genuine single photon as the walker and considering that the length of the crystal should be as short as possible to reduce dispersion and improve stability, the duration of the single photon pulse should be as small as possible. It is selected on the level of hundreds of femtoseconds. Such a short single photon pulse can be generated via SPDC with an ultra-short femtosecond pulse laser as the pumper. The generated photon pairs are time correlated, the click of detection on the idler photon can predict the existence of the signal photon. Various architectures exist for generating this type of heralded single photon from SPDC. Here, considering the features of high brightness and collection efficiency, we adopt the beam-like SPDC\,\cite{Huang2011}.

\subsection{Frequency up conversion single photon detection}
The spectrum of the arriving time of single photon is usually measured with the technology of time correlated single photon counting and commercial single photon detectors\,\cite{OConnor2012}. However, in our case, the signals are contained in a single photon pulse train with a pulse duration of approximately 1\,\emph{ps} and a repetition of 5\,\emph{ps}. Counting and analyzing such ultra-fast single photon signals are challenging. The time resolution of commercial single photon detectors is limited by the time jitter, which is typically in the range of tens to hundreds picoseconds\,\cite{Hadfield2009}. That is to say, it is unsuitable to directly use any commercial single photon detectors in our experiment. The detection of single photon with high resolution in time can be realized by transforming the temporal resolution to a spatial resolution. The measurement of an ultra-fast pulse of single photons can be realized via optical auto-correlation\,\cite{Trebino2012}, a technology developed from the optical parameter up conversion. That is, using an ultra-fast laser pulse to pump a nonlinear crystal, when the single photon and pumper pulse meet each other inside the crystal, the single photon will be up-converted to a shorter wavelength via the sum frequency process. For the photon with a long wavelength can be converted to a short one, this technology has been widely used in quantum communication for improving the detection efficiency in the infrared waveband\,\cite{Ma2012}. Here, we adopt this technology for its high resolution in time. Although periodically poled crystals are widely used in this technology for their high conversion efficiency, they are useless in our experiment for concentrating on the time resolution. The thickness of nonlinear crystal should be as thin as possible meanwhile taking into account the conversion efficiency. There exist two types of structures, collinear and non-collinear sum frequency. We adopt the latter to obtain a better signal to noise ratio (SNR), which is induced by the spatial divergence between the sum frequency signal and the pump laser. The crystal used in our experiment is a 1\,\emph{mm} thick $\beta$-BaB$_2$O$_4$ (BBO) crystal, cut for type-\uppercase\expandafter{\romannumeral2} second harmonic generation in a beam-like form. Then, the incidence angle of the signal pulse train and the pump laser are equal to each other, with $3^\circ$ to the normal direction. For reducing the noise induced by the strong pump laser, a dispersion prism in a 4F system is adopted as a spectrum filter. The scattered photons with wavelength longer than 395\,\emph{nm} are blocked by a knife edge. The rising edge in the sideband of this self-established spectrum filter is less than 1\,\emph{nm}.

\subsection{Detail description of the experimental setup}
An ultra fast pulse (140\,\emph{fs}) train generated by a mode-locked Ti:sapphire laser with a central wavelength at 800\,\emph{nm} and repetition ratio 76\,\emph{MHz} is firstly focused by lens L1 to shine on a 2\,\emph{mm} thick $\beta$-BaB$_2$O$_4$ crystal\,(BBO1), cut for type-\uppercase\expandafter{\romannumeral1} second harmonic generation. The frequency-doubled ultraviolet pulse (with a wavelength centered at 400\,\emph{nm}, 100\,\emph{mW} average power and horizontally polarized) and the residual pump laser are collimated by lens L2, and then separated by a dichroic mirror (DM). The frequency-doubled pulse train is then focused by lens L3 to pump the second nonlinear crystal\,(BBO2), cut for type-\uppercase\expandafter{\romannumeral2} non-degenerate beam-like SPDC. The signal and idler photons are collimated together with one lens (f=150\,\emph{mm}). The collimated signal photons in horizontal polarization with a center wavelength at 780\,\emph{nm} are then guided directly in free space to the following QW device. The collimated idler photons in vertical polarization with a center wavelength of approximately 821\,\emph{nm} firstly pass through a spectrum filter with a central wavelength 820\,\emph{nm} and bandwidth 12\,\emph{nm} and then are coupled into a single-mode fibre and sent directly to a single-photon avalanche diode for counting in coincidence with the signal photons. The quantum walks device is composed of HWPs and calcite crystals, and each step contains one piece of HWP and one piece of calcite crystal. In the experiment, we have adopted 50 such sets. The initial state is prepared by an apparatus composed of a PBS, a HWP and a QWP orderly. A reference laser beam for calibration is coupled into the QW device with this PBS. The residual pump in the frequency-double process is split by a PBS (not shown) into two beams, one acting as the reference laser and the other with most of the residual power adopted as the pump in the following frequency up conversion single photon detection. The partition of their power is realized by a HWP (not shown) and both of them are delayed with retroreflectors for matching the arriving time of the signal photons. After the quantum walks is finished, the signal photons are collected into a short single mode fibre (10\,\emph{cm} long) by a fibre collimator and then guided to the polarization analyzer composed of QWP, HWP and PBS orderly. Finally, the arriving time of signal photons is measured by scanning the pump laser and detecting the up conversion signals with a photomultiplier tubes. For reducing the scattering noise, BBO3 is cut for non-collinear up conversion and a spectrum filter based on a 4F system is constructed, where a prism is adopted for introducing the dispersion, a knife edge is used to block the long waves and the signal is reflected to the PMT with a pickup mirror. The inset in the bottom right corner gives the diagram of the time-multiplexing split-step quantum walks using birefringence crystals. The sites are defined as the arriving times of the photons. For each site, the photons are located within a pulse with a duration of a few hundred femtoseconds. For each split step, the horizontal photons will travel approximately 5\,\emph{ps} faster than the vertical ones for the birefringence in the calcite crystals, equivalently, the walker jumps to the right neighbouring site when the coin state is in horizontal. For a complete step, the original point is redefined, which results in the jump to the left neighbouring site when the coin state is vertical. The repetition rate of our laser is 76\,\emph{MHz}, which corresponds to a time interval $\sim$13\,\emph{ns}, significantly large than the total length of the lattice of approximately 0.25\,\emph{ns}.

\section{Full reconstruction of the final wave-function}

\begin{figure*}
  \centering
  \includegraphics[width=0.47\textwidth]{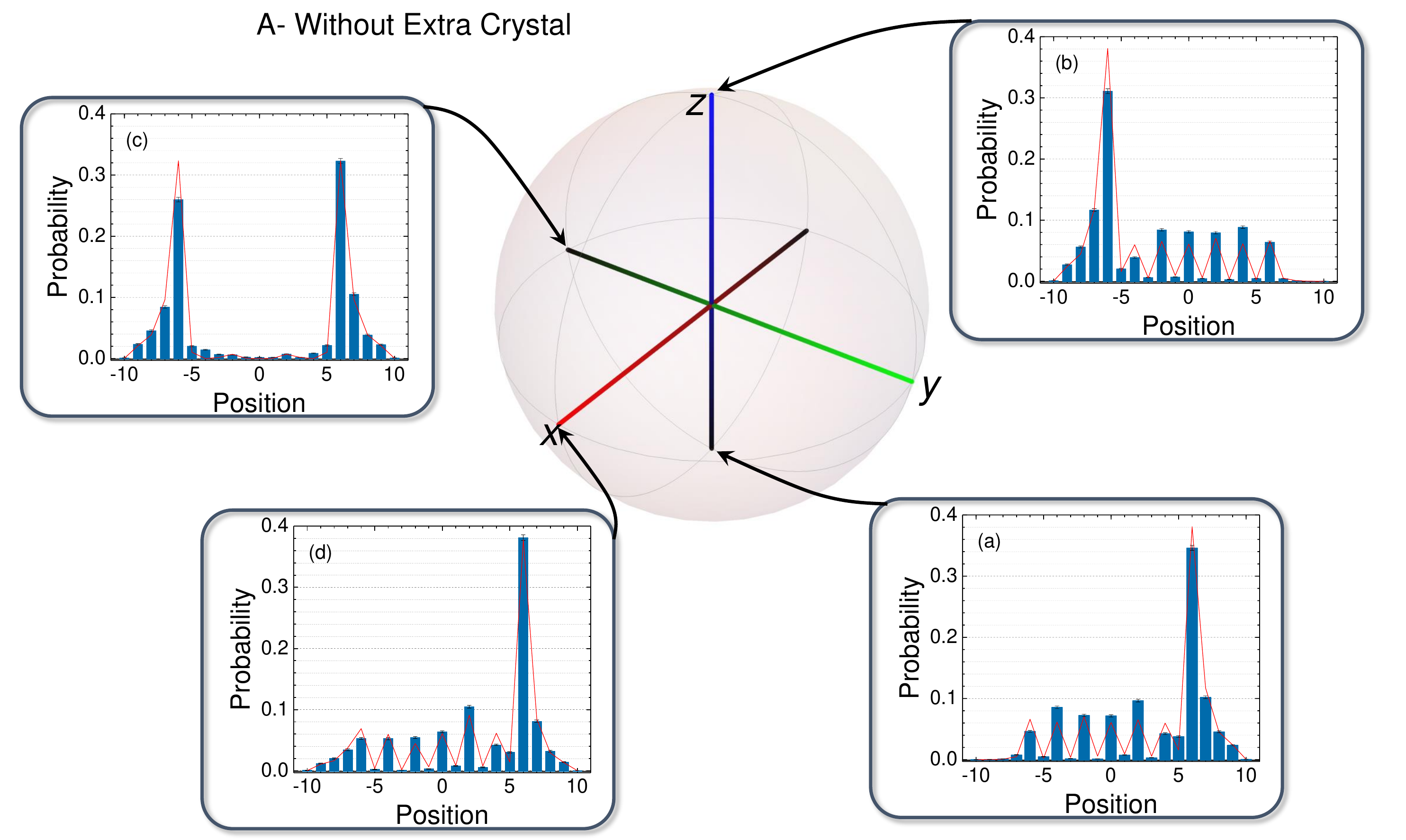}
  \includegraphics[width=0.47\textwidth]{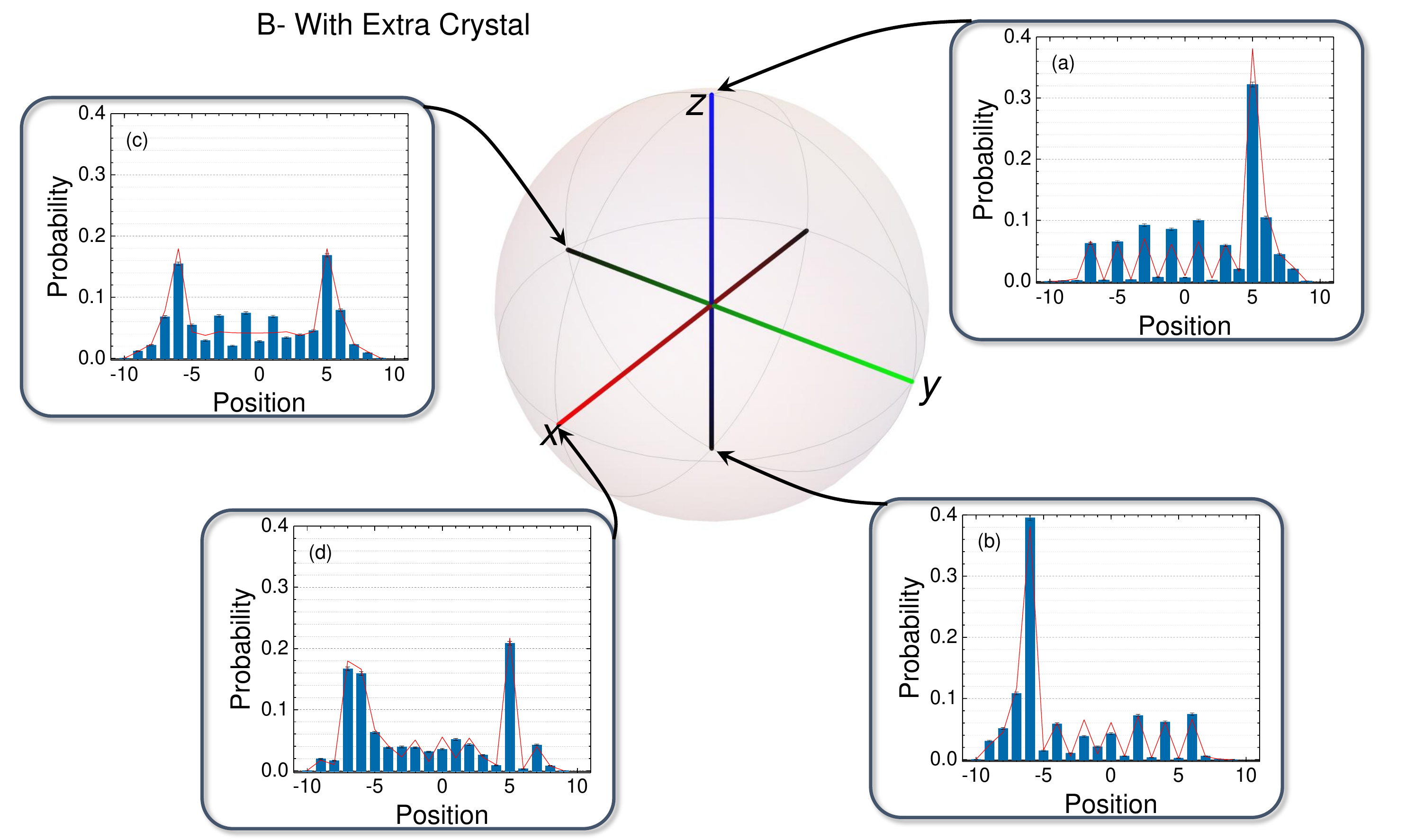}
  \caption{Distributions after projecting to the bases given in Eq.\ref{Eq.basises} for full wave-function reconstruction. A, before inserting the extra crystal. B, after inserting the extra crystal. In both of A and B, the spin is projected to four bases, (a) for $s^H$, (b) for $s^V$, (c) for $s^R$ and (d) for $s^D$, which corresponds to the eigenstates of the Pauli matrices, $\sigma_z = \pm 1, \sigma_y = -1$ and $\sigma_x = +1$. Experimental results are presented with histograms and the red lines gives the ideally expected values. The errors are given with considering the statistical noise.}\label{fig.BasesProb}
\end{figure*}

\subsection{Reconstruction of wave-function in site space}

We consider to experimentally reconstruct the full wave-function which was deemed to a challenge in usual interferometer based QWs\,\cite{Cardano2015}. Suppose the walker starts at the origin and consider the walker state at a certain step $t$,
\begin{equation}
  |\Psi_t\rangle =\sum_{x=-N}^N p_t(x)e^{-i\phi_t(x)} |\psi_t(x)\rangle\otimes|x\rangle,
\end{equation}
where the position index $x\in[-N,N]$ (integer) and the lattice size is $2N + 1$. For each position $x$ there is a local normalized spinor state $\ket{\psi_t(x)}$ with a complex amplitude $p_t(x)e^{-i\phi_t(x)}$. For convenience, we write local spinor state in
\begin{equation}
  \ket{\psi_t(x)} = \cos\frac{\theta_t(x)}{2}|H\rangle + e^{i\delta_t(x)}\sin\frac{\theta_t(x)}{2}\ket{V},
\end{equation}
with $\theta_t(x)\in[0,\pi]$ and $\delta_t(x)\in[0,2\pi)$. In experiment, we perform three steps to obtain the parameters, $p_t(x), \phi_t(x), \theta_t(x)$ and $\delta_t(x)$. Noting that the first lattice site phase $\phi_t(-N)$ is meaningless and the normalization condition $\sum_{-N}^{N}|p_t(x)|^2 \equiv 1$ should be satisfied. For each position, there exist four independent variables. Therefore there are $4(2N + 1)-2$ independent in total after a $t$-step walks starting from the original position ($N = t$).

Firstly, for each site $x$, we perform local projection measurement on the conventional bases
\begin{eqnarray} \label{Eq.basises}
% \nonumber % Remove numbering (before each equation)
  s^H &=& \ket{H}\bra{H}, \nonumber\\
  s^V &=& \ket{V}\bra{V}, \nonumber\\
  s^R &=& \ket{R}\bra{R}, \nonumber\\
  s^D &=& \ket{D}\bra{D},
\end{eqnarray}
where $\ket{R} = \frac{1}{\sqrt{2}}(\ket{H} - i\ket{V}), \ket{D} = \frac{1}{\sqrt{2}}(\ket{H} + \ket{V})$. We note the corresponding expected counts $n^H(x)$, $n^V(x)$, $n^R(x)$, and $n^D(x)$ ($x\in[-N,N]$). We can obtain a set of $4(2N+1)$ counts which is labeled by $\mathcal{S}$.

Secondly, an extra crystal is inserted and a spin echo is performed to shift all the horizontal bins a step backward. As a result, the spinor state at site $x$ is changed to be
\begin{equation}
  \ket{\psi'_t(x)} = \mathcal{N}\binom{p_t(x+1)e^{-i\phi_t(x+1)}\cos\frac{\theta_t(x+1)}{2}}{p_t(x)e^{i[-\phi_t(x)+\delta_t(x)]}\sin\frac{\theta_t(x)}{2}},
\end{equation}
where $\mathcal{N}$ is a renormalization coefficient. Then again we perform local projection measurement on the same bases and note the corresponding expected counts $\tilde{n}^H(x)$, $\tilde{n}^V(x)$, $\tilde{n}^R(x)$, and $\tilde{n}^D(x)$ ($x\in[-N,N]$). At this stage, we can obtain a set of $4(2N-1)$ counts which is labeled by $\tilde{\mathcal{S}}$.

At last, we carry out a numerical global optimization program based on simulated annealing algorithm to find the optimal pure state $\ket{\Psi_t}$ which can give the data set $\mathcal{S} + \tilde{\mathcal{S}}$. %\counts $n^H(x)$, $n^V(x)$, $n^R(x)$, $n^D(x)$ and $\tilde{n}^H(x)$, $\tilde{n}^V(x)$, $\tilde{n}^R(x)$, $\tilde{n}^D(x)$.
Following the method frequently used in measurement of qubits, we optimize the pure state $\ket{\Psi_t}$ by finding the minimum of the following ``likelihood" function\,\cite{James2001},
\begin{equation}
    \mathcal{L} = \sum_{x=-N}^N\sum_{i\in\{H,V,R,D\}}\frac{[\mathcal{N}n^i(x)-n^i_{\text{exp}}(x)]^2}{2\mathcal{N}n^i(x)} +  \sum_{x=-N}^{N-1}\sum_{i\in\{H,V,R,D\}}\frac{[\mathcal{N}\tilde{n}^i(x)-\tilde{n}^i_{\text{exp}}(x)]^2}{2\mathcal{N}\tilde{n}^i(x)},
\end{equation}
where $n^i_\text{exp}(x)~(\tilde{n}^i_\text{exp}(x))$ stands for the experimentally measured counts and $\mathcal{N}$ is a normalization coefficient. The total number of bases we have measured is $4(2N+1) + 4(2N-1)$, which is sufficiently large to reconstruct the final state with the assumption that the system is in a pure state.

\subsection{Obtaining the wave-function in quasi-momentum space}

The wave-function in quasi-momentum space $\ket{\Phi_t}$ corresponds to the Fourier transformation of the wave-function in site space. To obtain $\ket{\Phi_t}$, we perform a discrete Fourier transform to the reconstructed wave-function $\ket{\Psi_t}$. In details, $\ket{\Psi_t}$ contains two components, the complex amplitude $p_t(x)e^{i\varphi_t(x)}\cos{\frac{\theta_t(x)}{2}}$ for horizontal polarization and the complex amplitude $p_t(x)e^{i(-\varphi_t(x)+\delta_t(x))}\sin{\frac{\theta_t(x)}{2}}$ for vertical polarization. Two individual discrete Fourier transforms are performed to the two components. Then with a normalization for each quasi-momentum $k$, we can get the normalized spinor state $\phi_t(k)$ for each $k$. Theoretically, the time evolution operator $U(\theta_1,\theta_2)$ is diagonalized in Fourier basis, the final sate of spin for each quasi-momentum $k$, i.e., $|\phi_t(k)\rangle$ can then be obtained directly by performing a unitary operation (actually a rotation) on the initial state according to the time evolution operator.

%Supposing that the initial state $|\psi\rangle = \frac{1}{\sqrt{2}}(|\uparrow\rangle + i|\downarrow\rangle)$, the final states in quasi-momentum space after 500 steps are depicted on the Bloch sphere and dynamically shown in part\,(e) of the animations for explaining the topological phenomenons (Fig.\,SA4). In the topological trivial phase, the spinor final states in quasi-momentum space are localized to a certain region on the Bloch sphere in this phase. While for the non-trial phase, these final states can cover the full Bloch sphere.

% that the walker starts from position zero ($|0\rangle$) and the spinor state of the walker is initialized in a normalized spin sate $|\psi\rangle$. Then, in quasi-momentum space, the initial spinor state for each quasi-momentum $k$ is also $|\phi(k)\rangle_0 = |\psi\rangle$.

\begin{figure}
  \centering
  % Requires \usepackage{graphicx}
  \includegraphics[width=4in]{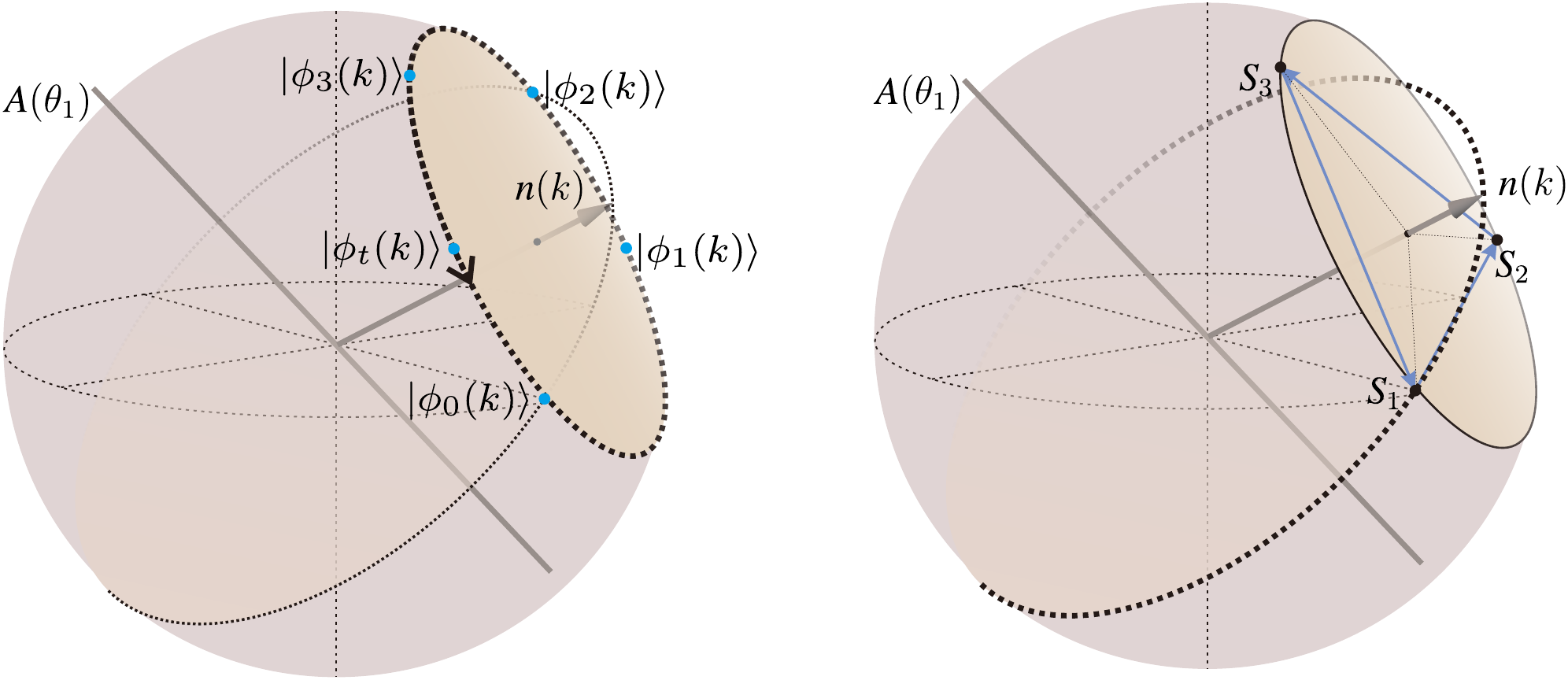}
  \caption{Diagram of the spinor state rotation in $k-$space (left sphere) and the reconstruction of eigenvectors $\bm{n}(k)$ from the spinor states (right sphere). The initial state $|\phi(k)\rangle_0$ will be rotated alone the vector $\bm{n}(k)$ after a certain step of quantum walk. As a result, the states for different steps $1, 2, 3, ... , t$ will be constrained to lie on a plane determined by $\bm{n}(k)$. For the spinor states after various steps of quantum walks are constrained to lie on a plane determined by $\bm{n}(k)$, generally, three individual points (black points, $S_1, S_2, S_3$) are enough for obtaining the normal vector of that plane.}\label{Fig.S1}
\end{figure}

\subsection{Reconstruction of eigenvectors}
Theoretically, the spinor states $|\phi_t(k)\rangle$ ($t=0,1,2,...$) for a fixed quasi-momentum $k$ after a $t$-step quantum walk, can be obtained by rotating the initial state $|\phi_0(k)\rangle$ with the angle $t.E(k)$ around the axis $\bm{n}(k)$, as sketched in the left panel of Fig.\,\ref{Fig.S1}. In other words, the spinor states $|\phi_t(k)\rangle$ will be constrained to lie on a plane that is perpendicular to $\bm{n}(k)$ on the Block sphere. As a result, there exists correspondence between the plane determined by the spinor states $|\phi_t(k)\rangle$ and spinor eigenvectors $\bm{n}(k)$. To determine the spinor eigenvectors $\bm{n}(k)$, generally, three different steps are enough to determine a plane (for the special cases that the vectors are linear dependent, the steps need more), sketched in right panel of Fig.\ref{Fig.S1}. Using this method, the sign of $\bm{n}(k)$ (`plus' or `minus', corresponding to two spin eigenvectors) remains uncertain. Resorting to the continuation of $\bm{n}(k)$ in $k$ space and assuming the direction of the first eigenvector $\bm{n}(k_0)$ (where $k_0$ can be chosen to be $-\pi$ or $0$) is fixed; the entire $\bm{n}(k)$ can then be uniquely determined. It should be noted that there exists a null set where the energy band is flat ($\theta_1 = \pm\pi$ or $\theta_2 = \pm\pi$). For the case $\theta_2 = \pm\pi$, located in the trivial phase, the walker will stand at the origin all the time and no changes can be observed. For the case $\theta_1 =\pm\pi$, there only exist one different final states compared to the initial state, therefore it will be failure to determine the plane and $\bm{n}(k)$ with only two points. While for other values of the parameters $\theta_1$ and $\theta_2$, our method is valid.

\section{Reading the topological phase from the reconstructed spinor eigenvectors}
\subsection{Standard time frame}

The topology in split-step DTQWs is firstly investigated in a standard time frame\,\cite{Kitagawa2010a}, that is,
\begin{equation}\label{Ustandard}
    U(\theta_1,\theta_2) = T_-R(\theta_2)T_+R(\theta_1).
\end{equation}
In this scenario, the eigenvectors $\bm{n}(k)$ are constrained to lie on a plane defined by the vector $\bm{A}(\theta_1)$, and as a result, the chiral symmetry can be defined with $\bm{A}(\theta_1)$. Although the winding numbers in spit-step DTQWs can be understood by the effective $\bm{n}(k)$ winding around $\bm{A}(\theta_1)$ on the Bloch sphere, directly getting these eigenstates is full of challenge. Alternatives have been developed to detecting the topological phase indirectly. The edge states have been observed\,\cite{Kitagawa2012b}, and the phase transitions between them are identified by the statistical moments of the final distributions\,\cite{Cardano2016}. Recently, they improved their method, the Zak phase connected to the winding number of the bulk of this system has been experimentally exploited by measuring the so called mean chiral displacement\,\cite{Cardano2017}. In our experiment, with the ability of full reconstruction of the final spinor states in $k$-space, we can read the winding number of $\bm n(k)$.
%, we can firstly reveal the different state evolution features in different topology, that is the global or local covering feature of the final spinor states. This covering feature gives a qualitative indicator of the topology, similar to the method of measuring the statical moments. Here for a quantitative measurement of the topological phase, we propose and experimentally demonstrate that the spinor states in the $k$-space can be obtained for each step in our quantum walk platform; and the $\bm{n}(k)$ for each $k$ can be further determined from the spinor states after different step of walks starting from the same single lattice site. The method has been explained in detail in the previous section. With the reconstructed eigenstate $\bm n(k)$, the winding number
% or the topological invariants, the Zak phase
%then can be directly calculated according its definition.
%\textbf{The Zak phase is defined as}
%\begin{equation}\label{ZakPhase}
%\gamma=\frac{1}{2}\int_{-\pi}^{\pi}\mathrm{d} k(\bm n\times\frac{\partial\bm n}{\partial k})\cdot\bm v_\Gamma
%\end{equation}
%\begin{equation}\label{ZakPhase1}
%\gamma=i\int_{-\pi}^{\pi}\mathrm{d} k\langle\bm n(k)|\partial_k|\bm n(k+1)\rangle
%\end{equation}

\subsection{Shifted time frame}
Further works show that the topological phase in periodically driven system is much more complicated and its complete classification should be modified with two bulk invariants\,\cite{Asboth2012,Asboth2013,Asboth2014,Obuse2015,Cedzich2016a,Cedzich2016b}. It is suggested to introduce nonequivalent shifted time frames, both of which maintain the chiral symmetry, to fully determine the topological phase\cite{Asboth2013,Obuse2015}. The time evolution operators are given by
\begin{eqnarray}\label{Ushifted}
    U'(\theta_1,\theta_2) &=& R(\theta_1/2)T_-R(\theta_2)T_+R(\theta_1/2),\\
    U''(\theta_1,\theta_2) &=& R(\theta_2/2)T_+R(\theta_1)T_-R(\theta_2/2)
\end{eqnarray}
The topological phases are then determined by the combined invariants $(\nu_0,~\nu_\pi)$, where $\nu_0 = (\nu'+\nu'')/2$ and $(\nu'-\nu'')/2$. $\nu'$ and $\nu''$ are conventional winding numbers defined through the Berry phase for $U'$ and $U''$ respectively. For $U'$ and $U''$ are identical only by switching $\theta_1$ and $\theta_2$, what we measure in experiment is $\nu'$, which is theoretically given by
\begin{eqnarray}
    \text{when}&~\sin^2(\theta_1)-\sin^2(\theta_2)>0,~~~~&
    \nu' = \begin{dcases}
               1  & (0<\theta_1<\pi)\\
              -1 & (-\pi<\theta_1<0)
          \end{dcases}\\
    \text{else,}& &\nu' =0
\end{eqnarray}

\subsection{Figures for discussion of robustness}
\begin{figure}
  \centering
  % Requires \usepackage{graphicx}
  \includegraphics[width=0.9\textwidth]{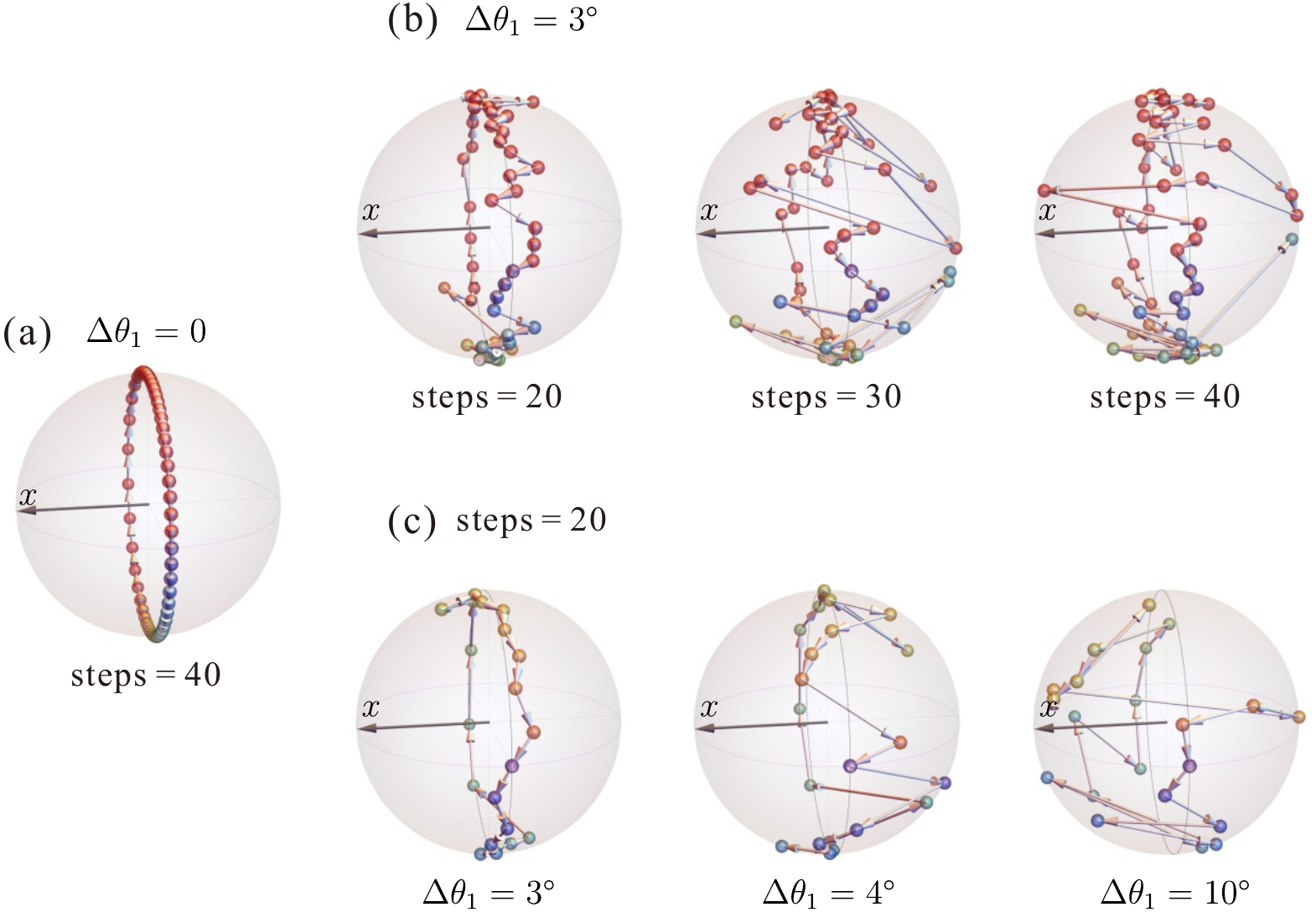}
  \caption{Numerical simulations of the winding of eigenvectors in the presence of dynamic disorder. Without loss of generality, the disorder is introduced via fluctuations of the first rotation angle $\theta_1$ over evolution time, the second rotation angle is $\theta_2 = 10^\circ$ and choose a mean value of the first rotation angle as $\bar\theta_1 = -22.5^\circ$. With the given parameters $(-22.5^\circ,10^\circ)$, the system is expected to yield a topological phase with its winding number $\mathcal{W} = -1$. For comparison, (a) shows the theoretical eigenvectors reconstructed from the 0-, 1- and 40-step quantum walks in the absence of disorder. In (b), the eigenvectors reconstructed from the 0-, 1- and 20-(30-, 40-, from left to right) step quantum walks in the presence of the dynamic disorder with disorder strength $|\Delta\theta_1|=3^\circ$. It can be observed that in this scenario, the winding of the eigenvector is robust when the number of the step is small, no matter the divergence of the eigenvectors to their ideal expectations (dashed circle). When the time gets larger, the decoherence induced by the dynamic disorder will diverge the eigenvector to reach the $x-z$ plane, which results in the failure of reading out the winding number. In (c), the eigenvectors reconstructed from the 0-, 1- and 20-step quantum walks with three different disorder strength $\Delta\theta_1 = 3^\circ$, $4^\circ$, $10^\circ$. We can see that the winding of the eigenvector is robust provided that the strength of the disorder is small (the energy gap is larger than noise spectral bandwidth). With increasing the disorder strength, reading out the eigenvector will be failed for the divergence of the eigenvector reaches the $x-z$ plane, as shown in the scenario $\Delta\theta_1 = 10^\circ$. The number of samples is 100 in (b) and (c).
  %The total number of steps is 20 and the chiral symmetry is defined by $\sigma_x$ (arrow) in shifted time frame. In (a), we set the mean value of parameters $\theta_1 = 22.5^\circ,~\theta_2 = 35^\circ$ (trivial phase $\mathcal{W} =0$) with a small random retardation $1^\circ,~2^\circ,~3^\circ$ at each step for $\theta_1$ (upper) and $\theta_2$ (lower). (b) and (c) correspond to the scenarios $\theta_1 = 22.5^\circ,~\theta_2 = 10^\circ$ with $\mathcal{W} = +1$ and $\theta_1 = -22.5^\circ,~\theta_2 = 10^\circ$ with $\mathcal{W} = -1$.
  }\label{Fig.Robust}
\end{figure}

\subsection{Statistical moments}
\begin{figure}
  \centering
  % Requires \usepackage{graphicx}
  \includegraphics[width=3.5in]{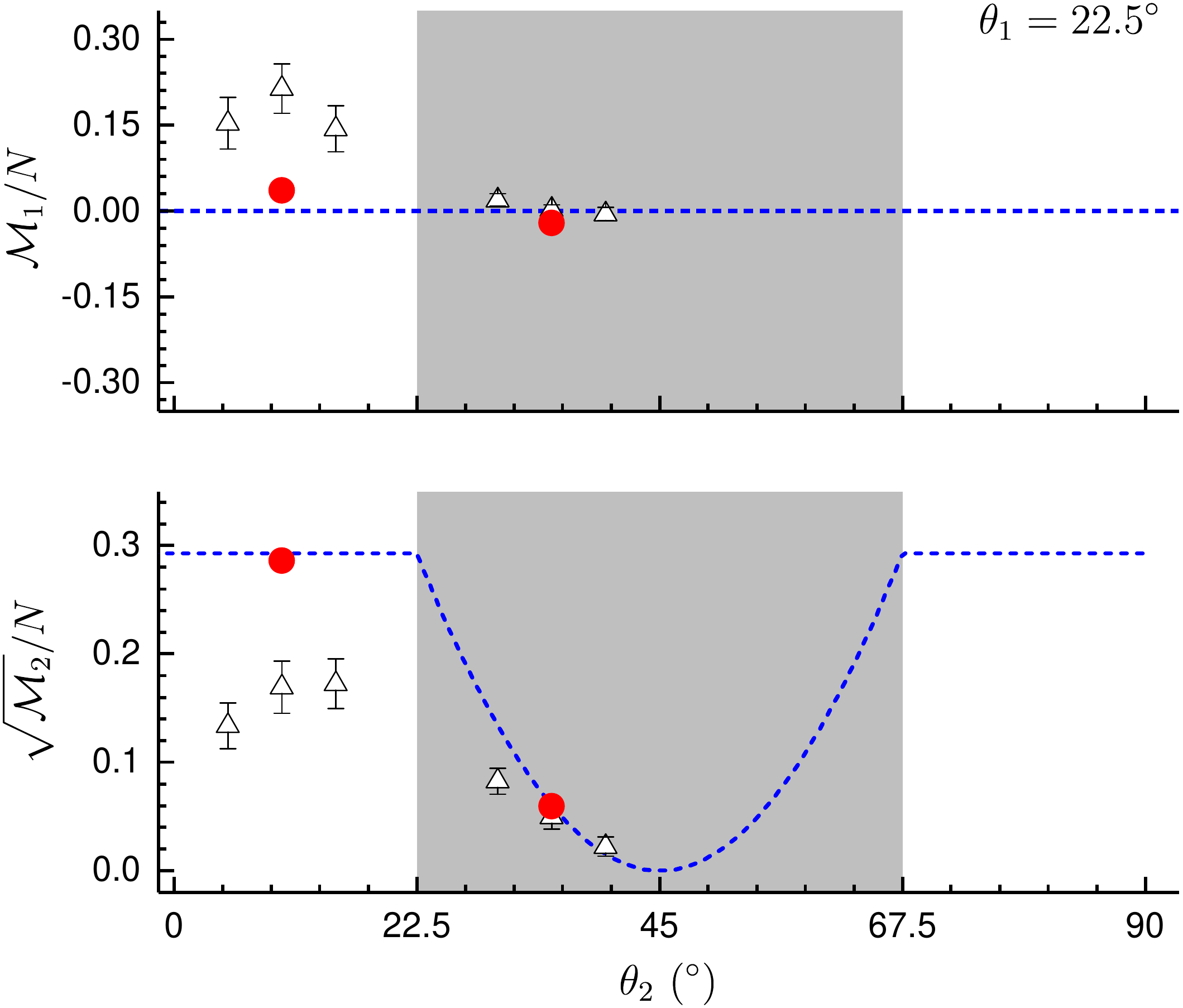}
  \caption{%Selected working points in phase diagram (a) and the measured statistical moments (b). Six points (three in topological non-trivial phase and three in topological trivial phase) are selected according the values of the second rotation angle $\theta_2$. Here for the tossing operation is realized by HWP, the range of rotation angles is within $[-90^\circ, 90^\circ]$ for both $\theta_1$ and $\theta_2$. The numbers in the legend give the corresponding winding number.
  Points show the experimentally measured normalized statistical moments after a 50-step QW (black triangles) and 20-step (red circles): first order (upper) and second order (lower) with the blue dashed lines representing the theoretical expectations. $\theta_1$ is fixed at $22.5^\circ$ with $\theta_2 = 5^\circ,~10^\circ,~15^\circ$ in non-trivial phase and $\theta_2 = 30^\circ,~35^\circ,~40^\circ$ in trivial phase for the 50-step scenario. In such large scale, the system is more sensitive and hard to measure. The decoherence will degenerate the QW's quantum feature, that is, experimentally measured $\mathcal{M}_2$ is lower than its theoretical predictions. In a smaller scale (20-step), the results match the theory very well (red circle). The region with shadow represents the trivial phase (non-trivial phase without shadow). We consider the statistical error and present the error bar with standard variance via numerical simulation.
  }\label{Fig.S4}
\end{figure}

The analytic forms of the statistical moments can be given in our scenario, which are
\begin{equation}\label{Eq.2moment}
    \mathcal{M}_2 = \tan^2{(\theta_1/2)}[1-\max{(|\sin(\theta_1/2)|,|\sin(\theta_2/2)|)}]
\end{equation}
for the second order moment and
\begin{equation}\label{Eq.1moment}
    \mathcal{M}_1 = \tan{(\theta_1/2)}[1-\max{(|\sin(\theta_1/2)|,|\sin(\theta_2/2)|)}][\langle\psi_0|(\sigma_x+\tan{(\theta_1/2)}\sigma_z)|\psi_0\rangle]
\end{equation}
for the first order moment (which is dependent on the initial state $\ket{\psi_0}$). In Fig.\ref{Fig.S4}, we show the measured statistical moments (1st and 2nd order) for 50-step and 20-step in different topological phases.

\section{Experimental imperfections, decoherence and noise.}
One of the main concerns of this work is to realize a large-scale DTQW with genuine single photons adopted as the walker. As a result, there are many challenges which usually can be neglected in DTQWs with small scale or attenuated laser as the walker.

\subsection{Blocking the scattering photons}
First of all, the walker is in the level of single photon, although the stray photons in environment are in extremely low level, the detection of signal adopts a high power laser (approximately 300\,\emph{mW}) as the pump, whose direction of propagation is the same as the walker, thus it is straightforward for the scattering photons from the strong pump to ultimately enter the single photon detectors. Here, we adopt three technologies to overcome this noise: Firstly, the SPDC for generating the heralded single photons is chosen to be non-degenerate in wavelength, then by inserting a spectrum filter before detecting the idler photon the weak reflection of the strong pump laser is blocked, and with the help of coincidence counting the noise can also be reduced. Secondly, the nonlinear crystal adopted for the parametric up conversion detector is cut in type-\uppercase\expandafter{\romannumeral2} and with a $3^\circ$ incidence angle for the signal and pump, the scattering photons from the pump are then filtered in the dimensions of both polarization and momentum. Thirdly, a spectrum filter based on 4F system with a dispersion prism inside is built up for blocking the scattering photons with a similar wavelength as the signal, which are generated from the weak nonlinear parametric process of the high power pump inside the crystal.

\subsection{Loss and detection efficiency}
Our scheme can overcome the extra loss that exists in the previous time multiplexing DTQWs, while the losses induced by the imperfect anti-reflection coating and the intrinsic absorption of the crystals should be considered in the case of a large-scale. Although the transmittance of a single crystal can be as high as 0.995, the total transmittance is reduced to 50\% after passing through two hundred surfaces and crystals with total length as long as half a meter. Another imperfection is the low detection efficiency in our parametric up conversion single photon detector. This imperfection arises from two reasons: the low detection efficiency (typical 20\%) for the photomultiplier tube and the low up conversion efficiency induced by the small nonlinear coefficient of the crystal and the pulse broadening induced by dispersion inside the crystal. The total measured detection efficiency can be as high as 0.5\% at last. Considering all of these disadvantages, the total coincidence counts is 6\,\emph{pairs/(s}$\cdot$\,\emph{mW)} with a pump power of 300\,\emph{mW} in the up conversion.

\subsection{Noise analysis}
\emph{Tolerance of optical axis-} Firstly we consider the noise from the poor orientation of the wave plates for realizing the tossing operation. The experimentally accessible orientation precision for each wave plate is typical $0.2^\circ$. We have supposed that the optical angles of the wave plates are oriented randomly in a range defined by a Possian distribution and have adopted the Monte Carlo simulations to evaluate the error. The numerical results show that this type of error is on the order of $10^{-5}$ for considering the fidelity.

\emph{Statistical noise-}
%The second source is the imperfection in the pulse overlap for each step. Here, considering the face that the visibility of the time domain Mach Zehnder interferometer formed by two crystals is above 0.998, we can suppose that this configuration will introduce a phase error on the level of $\lambda/750$ for each step. Although this type of error is induced by the imperfections of the time shift operators, it can be treated as imperfections in the tossing operators that introduce a phase error when performing tossing. Similar to the previous numerical simulation, the evaluated error is again extremely small.
Another important noise is the shot noise, which is also estimated by the Monte Carlo simulations. In our experiment, the total counts is around $2\times10^4$, which will introduce an error $(\sim 10^{-3})$ to the measured probability distribution and its fidelity, and the fidelity in full wave function reconstruction.
%In the experimental demonstration of the conversional DTQWs, the number of counts is low, which will introduce a considerable error ($\sim10^{-3}$). In the following experiment for investigating the topology, %the error induced by shot noise is relatively insignificant for adopting the attenuated laser as the walker.
%one thousand times Monte Carlo simulations were performed on our results for giving the confidence of reading out the winding number from the reconstructed eigenvectors. Besides the probability distribution, the site phase and the pure state in each site involved in the Fourier transform dependent on the reconstructed density matrix for each site, whose fidelity is significantly effected by the counts. Here for reducing the shot noise, we adopted the attenuated coherent state as the walker and the confidence is evaluated large than 0.978.
%In our numerical simulation, we fixed the counts for the sites with extremely low probability ($\sim10^{-3}$).In theory, there exist some sites, where the photon will appear with extremely low probability ($\sim10^{-4}$)}.}

\emph{Amplitude damping-}
Note that amplitude damping exists in our scheme for the loss mentioned above is weakly polarization dependent. Because the damping is systematic, we can overcome it through compensation. We have also checked the degeneration of the extinction ratio and interference visibility as the system increases in size, and they are consistent with the exponent rule, which indicates that based on the high visibility for single Mach Zehnder interferometer, these type of degeneration can be neglected.

\emph{Phase drifting-}
The collinear structure in the interferometer induces the extreme stability for single step. However, as the system size increases, the stability degenerates in experiment. The vibration is typically responsible for this stability degeneration in other bulk interferometers. %Here the experiment in a single trial is finished within only 2.5\,\emph{ns}, thus this vibration is not responsible for this degeneration as its typical frequency is on the order of kilohertz.
In our experiment, the vibration amplitude of the rotation stages (fixing the crystals) is only in the typical level of $\mu$rad, which will not responsible for the system's stability degeneration. The main governing factor is the temperature of crystals. We have monitored the drifting of the environment temperature and the stability of the DTQWs with 50 steps. The two patterns match each other very well in term of the period, approximate one thousand seconds, primarily because of the thermal expansion (typically on the order of $10^{-6}/^\circ$C). Although the refraction index also changes with the temperature at the same level, the birefringence is much more lower. To eliminate the influence of the temperature, we firstly reduce the temperature fluctuation to less than $0.5^\circ$C, and monitor the temperature as the data is collected.

\emph{Mode matching-}
Mode matching is unnecessary for the collinear structure in our scheme, while the imperfections of optical elements will introduce mode mismatching. We mainly consider the influences of two imperfections which will cause the decoherence: the first one is the length tolerance of birefringent crystals. Similar to the spatial shift interferometer, the length of the crystal determines the degree of the temporal shift. The tolerance of the length is $\pm 0.01$\,mm, implying that the corresponding phase tolerance is slightly greater than four waves. We fixed the crystals sequentially by carefully tilting the crystal around its optical axis that is perpendicular to the table to find the maximal interference visibility with a broadband laser. The same method is also used for locating the destructive interference in experiment. Another imperfection is that of the cut angle of the optical axis, which is supposed to be parallel to the surface for introducing a pure time shift. An imperfection in the cut angle of the optical axis will introduce a walk-off effect for the extraordinary ray. Considering a typical cut angle tolerance of $0.25^\circ$, the two spots for the two perpendicular polarizations will be displaced by approximately 70\,$\mu m$, which is large relative to the spot size ~1\,\emph{mm}. To overcome this decoherence, we carefully tilted the pitch angle of crystals to find the maximal visibility and then used a short single mode fibre (0.1\,\emph{m} long) as a spatial filter. Consequently, the spatial mode mismatching is not responsible for the decoherence in our case. Although we have carefully compensated the length tolerance of the crystals, limited by the interference visibility, the temporal mismatching is still the main contributor of decoherence in our scheme.

\end{document}